\UseRawInputEncoding

\documentclass[%
 reprint,
superscriptaddress,
pra,
]{revtex4-2}

\usepackage{dsfont}
\usepackage{svg}
\usepackage{graphicx}% Include figure files
\usepackage{dcolumn}% Align table columns on decimal point
\usepackage{physics}
\usepackage{amsmath}
\usepackage{bbold}
\usepackage{float}

\usepackage{hyperref}
\hypersetup{
    colorlinks=true,
	citecolor=blue,
    linkcolor=red,
	urlcolor = blue
}

\usepackage{bm}% bold math
\usepackage[mathlines]{lineno}% 

\newcommand{\tauL}{\tau^{\mathcal{L}}_{m}}

\definecolor{LighterBlue}{rgb}{0.18, 0.48, 0.88}

\begin{document}
%\title{Non-Markovian Mpemba effect in a Quantum System}% 
%\title{Calculating two time correlations using dynamical maps}
\title{Accelerated calculation of impurity Green's functions exploiting the extreme Mpemba effect}
\author{David J. Strachan}
\email{david.strachan@bristol.ac.uk}
\affiliation{\textit{H. H. Wills Physics Laboratory, University of Bristol, Bristol BS8 1TL, United Kingdom}}
\author{Archak Purkayastha}
\email{archak.p@phy.iith.ac.in}
\affiliation{\textrm{Department of Physics, Indian Institute of Technology, Hyderabad 502284, India}}
\author{Stephen R. Clark}
\email{stephen.clark@bristol.ac.uk}
\affiliation{\textit{H. H. Wills Physics Laboratory, University of Bristol, Bristol BS8 1TL, United Kingdom}}

\date{\today}   
\begin{abstract}
Simulating the dynamics of quantum impurity models remains a fundamental challenge due to the complex memory effects that arise from system-environment interactions. Of particular interest are two-time correlation functions of an impurity, which are central to the characterization of these many-body systems, and are a cornerstone of the description of correlated materials in dynamical mean field theory (DMFT). In this work, we extend our previous work on the extrapolation of single-time observables to demonstrate an efficient scheme for computing two-time impurity correlation functions, by combining the non-Markovian quantum Mpemba effect (NMQMpE) with a dynamical map-based framework for open quantum systems. Our method is benchmarked against exact and known accurate results in prototypical impurity models for both fermionic and bosonic environments, demonstrating significant computational savings compared to state-of-the-art methods.

\end{abstract}
\maketitle
\section{Introduction}
Real quantum systems are never perfectly isolated, but instead are embedded within a larger environment and interacting with its excitations over a range of energies. Computing the behaviour of {\em open quantum systems}, such as the evolution of system observables and two-time correlations of the system, is a formidable many-body problem due to growth of complex system-environment correlations. Yet this information is fundamental for understanding equilibrium properties, response functions, and experimental observables~\cite{rickayzen2013green} of such systems relevant to many fields, including quantum optics~\cite{10.1093/acprof:oso/9780199213900.001.0001}, chemical physics~\cite{rivas2012open}, quantum biology \cite{lambert2013quantum,van2000photosynthetic,may2023charge}, and quantum technologies \cite{nielsen2001quantum,josefsson2018quantum,ronzani2018tunable,mosso2019thermal}. Quantum impurity models are a particular paradigmatic class of open quantum systems where the system of interest is very small, such as a single spin-1/2, yet they display a diverse spectrum of physical phenomena, including novel transport~\cite{PhysRevLett.96.216802,PhysRevLett.121.137702}, dissipation and non-equilibrium physics~\cite{PhysRevB.107.195101}. Originally developed to study the physics of magnetic impurities~\cite{PhysRev.124.41} in metals, in recent years they have gained additional relevance with the powerful dynamical mean field theory (DMFT)~\cite{RevModPhys.68.13}, where they provide the foundational ingredient for capturing strongly correlated effects in quantum materials. 

A key part of the DMFT loop is the calculation of a specific two-time correlation function, the impurity's single-particle Green's function. Due to the presence of non-trivial memory timescales, standard open quantum system approaches based on Markovian master equations are unsuitable to describe crucial impurity model physics. For this reason many techniques have been developed to overcome this, namely quantum Monte Carlo methods~\cite{PhysRevLett.93.136405,RevModPhys.83.349}, process tensor implementations of path integrals \cite{strathearn2018efficient,PhysRevB.107.195101,FEYNMAN1963118,PhysRevX.11.021040,Cygorek2022,Pollock2018,Fux2023,Cygorek2024a}, the hierarchical equations of motion \cite{10.1063/5.0011599} approach, perturbative projection operator techniques \cite{10.1063/1.1624830,10.1093/acprof:oso/9780199213900.001.0001,BREUER200136} to name just a few. Although successful, a major bottleneck in many methods is an unfavorable scaling of complexity with simulation time, preventing access to longer timescales. This is particularly an issue when thermalisation of the impurity is required so that the two-time correlation function of its equilibrium state can be computed. Improving the efficiency of this calculation is the focus of this work. 

Here we propose two steps based on a dynamical map approach which can significantly reduce the complexity of two-time correlation function calculations for impurity models. The first step is to exploit the recently discovered non-Markovian quantum Mpemba effect (NMQMpE) \cite{PhysRevLett.134.220403} to identify an initial state of the impurity that provides the fastest relaxation path to equilibration. Given any real-time dynamical solver for an impurity model this, by definition, is the least computationally expensive way to prepare a thermal state. The second step is to use the dynamical map to accurately extrapolate transient dynamics beyond the memory time of the environment up to stationarity for the system with minimal computational cost.
This extends our approach for extrapolating equal-time observables \cite{strachan2024extracting} to equilibrium two-time correlation functions. Our method is time-local but bares some similarities to the transfer tensor method~\cite{rosenbach2016efficient,PhysRevA.96.062122,doi:10.1021/acs.jpclett.6b02389,10.1063/1.5009086,WuCerrilloCao+2024+2575+2590} which is based on a discretised memory kernel approach and is thus time non-local, although recent work has introduced a new time-local transfer matrix scheme for extracting single time dynamics~\cite{cygorek2025timenonlocalversustimelocallongtime}. To date, the transfer tensor method has only been applied to bosonic environments thus far.

The structure of this paper is as follows: in Sec.~\ref{sec:nonmarkov} we outline the description of open quantum systems and equilibrium correlation functions using the time convolutionless master equation and dynamical maps, leading to the generalisation of the extrapolation we proposed in Ref.~\cite{strachan2024extracting} to two-time correlation functions. We then introduce the non-Markovian quantum Mpemba effect~\cite{PhysRevLett.134.220403} and present our full methodology to calculate equilibrium correlation functions, including how we calculate dynamical maps. In Sec~\ref{sec:Time evolution} we briefly introduce the thermofield chain mapping and present additional details of the implementation including the use of matrix product states (MPS), before presenting the results in Sec~\ref{sec:Results} for both fermionic and bosonic environments, including the paradigmatic single impurity Anderson model and spin-boson model. This is followed by our conclusions and outlook for future work in Sec~\ref{sec:Conclusion}.

\section{Non-Markovian Dynamics} \label{sec:nonmarkov}
We first introduce the time convolutionless equation in generality and it's associated memory timescales, before discussing our specific methodology. Throughout this work we take $k_{B} = 1 = \hbar$.

\textit{Open Systems} - 
Consider a system $S$ interacting with a large thermal environment $E$ with combined Hilbert space $\mathcal{H}_{SE} = \mathcal{H}_{S}\otimes\mathcal{H}_{E}$, with total Hamiltonian
\begin{equation} \label{eq:1}
\hat{H} = \hat{H}_{S}+\hat{H}_{SE}+\hat{H}_{E},
\end{equation}
where $\hat{H}_{S}$ is the Hamiltonian of the system, $\hat{H}_{E}$ is the Hamiltonian of the environment and $\hat{H}_{SE}$ is the coupling between them. Given an initial state of the full system $\hat{\rho}_{\rm tot}(0)$, the state of the system at a later time $t$ is given through the unitary evolution
\begin{equation} \label{eq:2}
\hat{\rho}(t)= \textrm{Tr}_{E}\big(e^{-i\hat{H}t}\hat{\rho}_{\textrm{tot}}(0)e^{i\hat{H}t}\big),
\end{equation}
which can be described by a dynamical map $\hat{\rho}(t) = \hat{\Lambda}(t)\hat{\rho}(0)$ if the initial state has the factorised form $\hat{\rho}_{\textrm{tot}}(0) = \hat{\rho}(0)\otimes \hat{\rho}_{E}$.
%$\dot{\rho}_{tot}(t) = -\frac{i}{\hbar}#[H,\rho_{\rm tot}(t)] \equiv \hat{\mathcal{L}}_{\rm tot}\rho_{tot}(t)$
As the environment contains many (possibly infinite) degrees of freedom and we are primarily interested in the system, we would like a reduced description. 

As derived in Ref.~\cite{10.1093/acprof:oso/9780199213900.001.0001}, the time convolutionless (TCL) master equation provides an exact, closed equation for the reduced dynamics, given by 
\begin{equation} \label{eq:3}
\frac{\textrm{d} \hat{\rho}(t)}{\textrm{d} t}  = \hat{\mathcal{L}}(t)\hat{\rho}(t) + \hat{\mathcal{I}}(t)\hat{\rho}_{\rm tot}(0), 
\end{equation}
where $\hat{\mathcal{L}}(t) = \frac{\rm d}{\textrm{d}t}[\hat{\Lambda}(t)]\hat{\Lambda}^{-1}(t)$ and $\hat{\mathcal{I}}(t)$ accounts for any initial system-environment correlations. We consider setups where the system approaches a unique steady state in the long-time limit, meaning $\lim_{t\to\infty}\hat{\rho}(t)=\hat{\rho}_\infty$ for any initial state $\hat{\rho}(0)$. Considering Eq.~\eqref{eq:3} in the long time limit gives
\begin{equation} \label{eq:4}
0= \hat{\mathcal{L}}(t)\hat{\rho}_{\infty} + \hat{\mathcal{I}}(t)\hat{\rho}_{\rm tot}(0).
\end{equation}
For $\hat{\rho}_\infty$ to be independent of $\hat{\rho}_{\rm tot}(0)$, we must have $\lim_{t\to\infty}\hat{\mathcal{I}}(t)\hat{\rho}_{\rm tot}(0)=0$, which then implies $\lim_{t\to\infty}\hat{\mathcal{L}}(t)\hat{\rho}_{\infty}=0$. This gives two timescales 
$\tau^{I}_{m}$, $\tauL$ given by
\begin{align} 
& ||\hat{\mathcal{L}}(t)[\hat{\rho}(\infty)]||<\epsilon~~ \forall~~ t \geq\tauL, \label{eq:5}\\
& ||\hat{\mathcal{I}}(t)[\hat{\rho}_{\rm tot}(0)]||<\epsilon~~ \forall~~ \tau \geq\tau_{m}^{I}, \label{eq:6} 
\end{align}
where $\epsilon$ is some arbitrarily small tolerance and $||\hat{A}||$ is the norm of $\hat{A}$. Here $\tauL$ defines the timescale over which a steady state can be well defined and $\tau_{m}^{I}$ defines the timescale at which the memory of initial system-environment correlations are lost. In Appendix~\ref{appendix: NZ equation} we provide a complementary discussion based on the Nakajima-Zwanzig equation. 

A key point here is that often $\tauL$ and $\tau_{m}^{I}$ are much less than the typical time for the system to relax to stationarity, such that given access to $\hat{\mathcal{L}}(t)$ we can extract the steady state with significantly less numerical resources than direct time evolution into the steady state. In addition, it also states that $\hat{\mathcal{L}}(t)$ can be used to describe the evolution of systems which do not have a factorised form at $t=0$, as long as $t>\tau_{m}^{I}$. Given one further assumption, we can extract full dynamics with minimal computational cost. If we assume the full propagator converges up to an error $O(\epsilon)$ on the timescale $\tauL$, so $\hat{\mathcal{L}}(t) = \hat{\mathcal{L}}_{\rm m}+O(\epsilon)$, $\forall \,t \geq \tauL$, then we have the following factorised form for Eq.~\eqref{eq:2}
\begin{equation} \label{eq:7}
\hat{\rho}(t) =  e^{(t-\tau^{*}_{m})\hat{\mathcal{L}}_{m}}\hat{\rho}(\tau^{*}_{m}) \; \forall~~t>\tau^{*}_{m}\geq\tauL,\tau_{m}^{I},
\end{equation}
where $\tau^{*}_{m}$ is taken to be whichever is the larger timescale out of $\tauL,\tau_{m}^{I}$.
This decomposition represents a phenomenon called {\em initial slippage} which has been widely studied \cite{Bruch_2021,GEIGENMULLER198341,PhysRevA.28.3606,PhysRevA.32.2462,Gaspard1999Slippage} and is closely associated with the assumption of a finite memory time for the environment correlations \cite{PhysRevX.11.021041,Bruch_2021}.

\textit{Equilibrium Correlation Functions} - In this work our focus is on the equilibrium two-time correlation functions $C_{A}(t) \equiv \langle \hat{A}(t)\hat{A}\rangle_{\infty}$ of some system operator $\hat{A}$ which describe how a system in equilibrium responds to a perturbation by $\hat{A}$, where $\hat{A}(t)$ is in the Heisenberg picture. Formally it is given by
\begin{align} \label{eq:8}
C_{A}(t) &= \textrm{Tr}(e^{i\hat{H}t}\hat{A}e^{-i\hat{H}t}\hat{A}\hat{\rho}_{\textrm{tot}}(0)), \nonumber \\
&=  \textrm{Tr}(\hat{A}e^{-i\hat{H}t}\hat{A}\hat{\rho}_{\textrm{tot}}(0)e^{i\hat{H}t}), \nonumber \\
&=  \textrm{Tr}_{S}(\hat{A}\tilde{\rho}_{A}(t)),
\end{align}
where we have introduced $\tilde{\rho}_{A}(t) = \textrm{Tr}_{E}\big(e^{-i\hat{H}t}\hat{A}\hat{\rho}_{\textrm{tot}}(0)e^{i\hat{H}t}\big)$. Here we take the system at time $t = 0$ as being in equilibrium with the environment, such that the total state $\hat{\rho}_{\textrm{tot}}(0)$ gives $\textrm{Tr}_{E}(\hat{\rho}_{\textrm{tot}}(0)) =\hat{\rho}_{\infty}$. The total state $\hat{\rho}_{\textrm{tot}}(0)$ possesses system-environment correlations and is taken as arising from the time-evolution from a time $t_0 <0$ of a factorised initial state $\hat{\rho}_{\textrm{tot}}(t_{0}) = \hat{\rho}(t_{0})\otimes\hat{\rho}_{\beta}$, in which the environment is assumed to be in a grand canonical Gibbs state
\begin{equation} \label{eq:9}
\rho_{\beta}=\frac{e^{-\beta(\hat{H}_{E}-\mu\hat{N}_{E})}}{\textrm{Tr}_{E}(e^{-\beta(\hat{H}_{E}-\mu\hat{N}_{E})})},
\end{equation}
where $\hat{N}_{E}$ is the number operator of the environment, $\mu$ is the chemical potential and $\beta$ is the inverse temperature. Once $t_{0} \to -\infty$ the system is thermalised at $t=0$. 

The two-time correlation $C_{A}(t)$ is fully described by the evolution of a perturbed density matrix $\tilde{\rho}_{A}(t)$, which is distinct from $\hat{\rho}(t)$ in two ways, (i) it does not satisfy the properties of a density matrix and (ii) it is not initially factorisable as $\tilde{\rho}_{\textrm{tot}, A}(0) = \tilde{\rho}_{A}(0)\otimes \hat{\rho}_{B}$. However, it undergoes the same time evolution as described by Eq.~\eqref{eq:2}, and is thus also described by the TCL master equation like $\hat{\rho}(t)$~\cite{10.1093/acprof:oso/9780199213900.001.0001,BREUER200136}. Note that if $\hat{A}$ is an odd parity fermionic operator, a parity correction is required which is detailed in Appendix~\ref{appendix: parity correction}. If both memory times $\tauL,\tau_{m}^{I}$ have been reached, meaning the memory of initial system-environment correlations are lost and $\hat{\mathcal{L}}(t)$ has become time-independent, then we have 
\begin{equation} \label{eq:10}
C_{A}(t) = \textrm{Tr}(\hat{A}e^{(t-\tau^{*}_{m})\hat{\mathcal{L}}_{m}}\tilde{\rho}_{A}(\tau^{*}_{m})), \; \forall~t>\tau^{*}_{m}\geq\tauL,\tau_{m}^{I},
\end{equation}
where again $\tau^{*}_{m}$ is taken to be whichever is the larger timescale of $\tauL,\tau_{m}^{I}$. This is the key result used in our approach.

\begin{figure}[t!] % t or b to control placement
  \centering
    \includegraphics[width=1\linewidth]{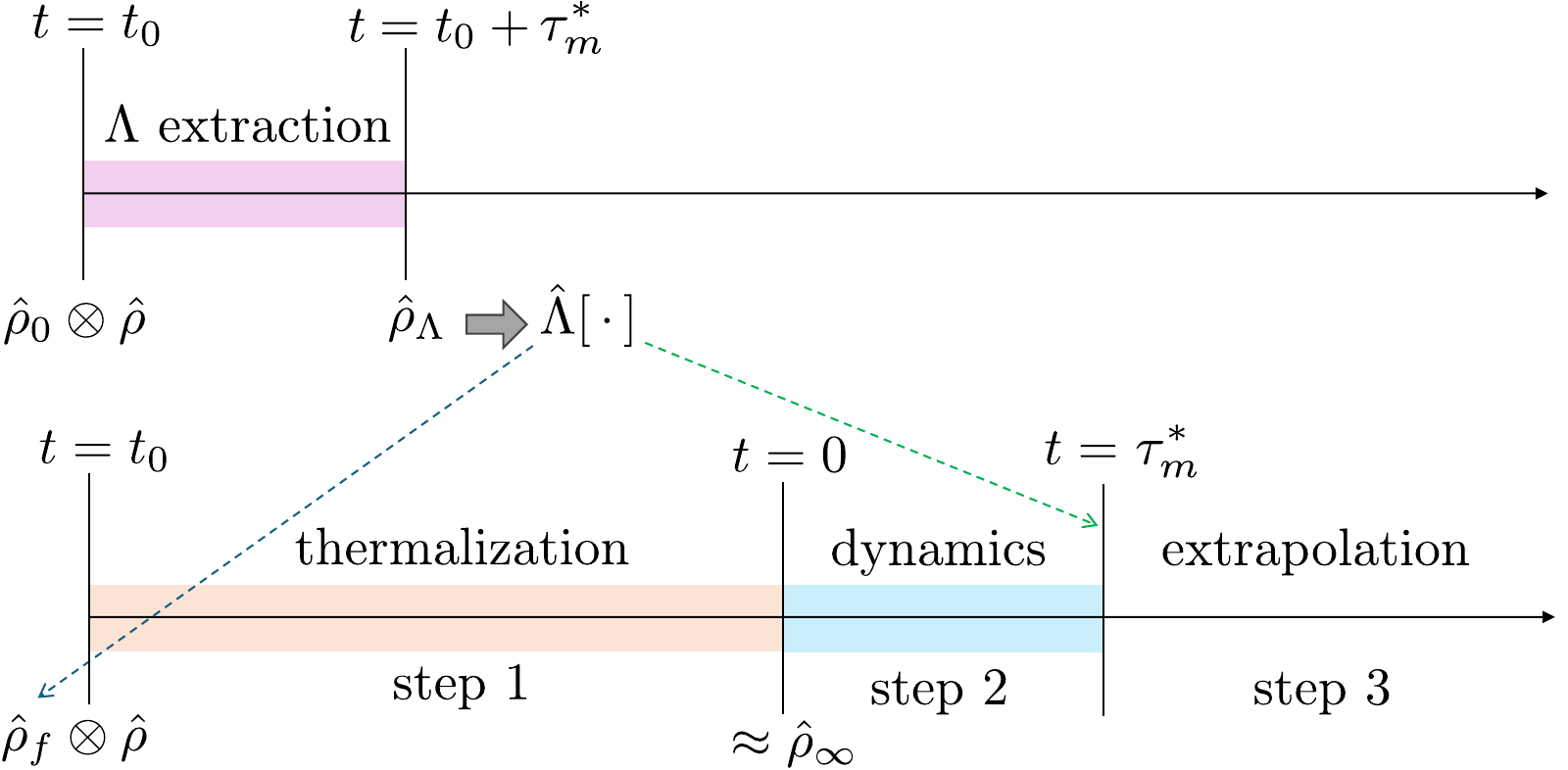}
    \caption{Schematic of our methodology. The top timeline represents the evolution of the Choi state, which allows for the extraction of the dynamical map $\hat{\Lambda}(t)$ as detailed in the Appendix~\ref{appendix:Extracting map}. This is needed for the calculation of $\hat{\rho}_{f}$ and the extrapolation. The bottom timeline represents the thermalisation from $t=t_0$ to $t=0$, at which point the system is perturbed by $\hat{A}$ and the combined system-environment setup undergoes unitary evolution until Eq.\eqref{eq:10} is valid, at which point extrapolation is used for the remaining time evolution.}
  \label{fig:schematic setup}
\end{figure}

\textit{Methodology} - The calculation of equilibrium correlation functions contains two time evolutions, the equilibration of $\hat{\rho}(t)$ and the dynamics of $\tilde{\rho}_{A}(t)$ which begins by perturbing $\hat{\rho}(t)$ once it has equilibriated. Both can be done using known methods, but are non-trivial many body calculations that are numerically costly. Our methodology provides significant speed ups for both parts of this calculation. First, we choose the initial state that equilibriates to $\hat{\rho}_{\infty}$ in the fastest possible time, which is not in general $\hat{\rho}_{\infty}$ itself, but rather
\begin{equation} \label{eq:11}
\hat{\rho}_{f} \equiv \frac{\hat{\Lambda}^{-1}(\tauL)\hat{\rho}_{\infty}}{\textrm{Tr}(\hat{\Lambda}^{-1}(\tauL)\hat{\rho}_{\infty})}.
\end{equation}
 This is the non-Markovian quantum Mpemba effect (NMQMpE)~\cite{PhysRevLett.134.220403} and can be understood from Eq.~\eqref{eq:7}. As discussed in Ref.~\cite{PhysRevLett.134.220403}, $\hat{\rho}_{f}$ is not guaranteed to be physical, particularly when $\hat{\rho}_{\infty}$ sits near the boundary of physical states, and it requires $\hat{\Lambda}^{-1}(\tauL)$ to be well-defined, which is not always the case. However, this only happens at isolated time points~\cite{Reimer2019Jul,PhysRevB.83.115416,li2012timeconvolutionless,chruscinski2010non}, such that we are free to shift $\tauL$ slightly to locate a well-defined $\hat{\Lambda}^{-1}(\tauL)$. For all the cases considered here $\hat{\rho}_{f}$ is physical, and only in the spin boson setup are we required to shift $\tauL$ to give a physical $\hat{\rho}_{f}$. Second, for the evolution of $\tilde{\rho}_{A}(t)$, once we reach the timescale such that Eq.~\eqref{eq:10} is valid we use the fixed $\hat{\mathcal{L}}_{m}$ to evolve $\tilde{\rho}_{A}(t)$ which for small systems is a trivial calculation. To summarise, 
\begin{enumerate}
  \item Unitarily evolve $\hat{\rho}_{f}\otimes\hat{\rho}_{\beta}$ up to $\tauL$ using a given time evolver. The system is then thermalised to $\hat{\rho}_{\infty}$.
  \item Apply the operator $\hat{A}$, and again unitary evolve the combined state of the system and environment up until $t>\tau_{m}^{I},\tauL$.
  \item Trace out the environment and evolve the perturbed reduced density matrix $\tilde{\rho}_{A}(t)$ using the converged propagator, $\hat{\mathcal{L}}_{m}$.
\end{enumerate}
This procedure is schematically shown in Fig.~\ref{fig:schematic setup}.

\textit{Extracting the Dynamical Map} - To implement our methodology, we need to extract the dynamical map $\hat{\Lambda}(t)$ and the time-local propagator $\hat{\mathcal{L}}(t)$. To do this we use the Choi-Jamiolkowski isomorphism which provides an equivalence between quantum states and superoperators~\cite{CHOI1975285,JAMIOLKOWSKI1972275}. By introducing ancillary replica modes for each system mode and maximally entangling them, we can fully reconstruct $\hat{\Lambda}(t)$ from a single pure state calculation of the unitary dynamics of the system, environment and their ancilla modes up to time $t$. For more details, see Appendix~\ref{appendix:Extracting map}. Instead, we could extract $\hat{\Lambda}(t)$ from $d$ calculations where $d$ is the dimension of $\mathcal{H}_{S}$, each evolving a basis state of the system as is done in Refs.~\cite{rosenbach2016efficient,PhysRevA.96.062122,doi:10.1021/acs.jpclett.6b02389,10.1063/1.5009086,cygorek2025timenonlocalversustimelocallongtime,WuCerrilloCao+2024+2575+2590}. For the models considered here, our approach which exploits the principle of superposition is a more efficient way to calculate $\hat{\Lambda}(t)$.

\section{Time Evolution} \label{sec:Time evolution}
To implement our approach the many-body time-evolution has to be computed in order to calculate the dynamical map, implementing rapid thermalisation (step 1) and the time evolution of the perturbed state $\tilde{\rho}_{A}(t)$ (step 2). Here we describe the specifics of our implementation. 

\textit{Chain Mapping} - Key challenges in simulating open systems include discretizing a continuous environment while minimizing finite-size effects, efficiently capturing the thermal state of the environment, which consists of a mixture of many states, and modeling the system’s coupling to the full environment without introducing long-ranged terms that drive excessive entanglement growth. To overcome this we apply the thermofield chain mapping~\cite{PhysRevA.92.052116,kohn2022quench} and use MPS techniques for time-evolution. First, a purification and change of single-particle basis leads to a trivial representation of the thermal state as a single Fock state, i.e. a pure state with no initial entanglement. The finite-size effects in the discretisation now manifest as excitations reflecting off the far boundary of the chain, and so long as this does not occur in the simulation time, the dynamics on the chain faithfully reproduce the effect of a continuous environment~\cite{10.1063/1.4940436}. Lastly, the coupling of the system to the environment now reduces to localised couplings to the ends of two chains, and the couplings within those chains are nearest-neighbour. All these features are well-suited to MPS techniques, resulting in this combination giving highly accurate results over moderate timescales. The exclusive use of chain-mapping and MPS provides a near-exact comparison to validate our acceleration steps.

More precisely, we consider a non-interacting environment with energy range $\omega \in \{\omega_{1},\omega_{2}\}$ described by the Hamiltonian $\hat{H}_{E}$ interacting with the system via the Hamiltonian $\hat{H}_{SE}$ 
\begin{equation} \label{eq:12}
\hat{H}_{E}+\hat{H}_{SE} = \int_{\omega_{1}}^{\omega_{2}}d\omega\omega\hat{a}^{\dag}_{\omega}\hat{a}_{\omega} + \int_{\omega_{1}}^{\omega_{2}}d\omega\sqrt{J(\omega)}(\hat{a}^{\dag}_{\omega}\hat{Q}+\hat{Q}^{\dag}\hat{a}_{\omega}),
\end{equation}
where $\hat{Q}$ is the system operator coupling to the environment, $J(\omega)$ is the environment spectral density, and $\hat{a}^{\dag}_{\omega},\hat{a}_{\omega}$ are canonical creation and annihilation operators obeying $[\hat{a}^{\dag}_{\omega},\hat{a}_{\omega'}]_{\pm}=\delta(\omega-\omega')$, taking $+$ for fermions and $-$ for bosons. After introducing an ancillary replica Hilbert space $\mathcal{H}_{A}$ for the environment and performing the thermofield transformation, the Gibbs state is mapped to a pure state $\hat{\rho}_{\beta}\to\ket{\Omega}$ such that $\textrm{Tr}_{A}(\ket{\Omega}\bra{\Omega}) = \hat{\rho}_{\beta}$. The two rotated environments are then mapped to chain geometries using a transformation based on orthogonal polynomials which leads to the following total Hamiltonian for the environment

\begin{align} \label{eq:13}
\hat{H}_{A}+ \hat{H}_{E}+\hat{H}_{SE} &= \sum_{i=1}^{2}\bigg[g_{i,0}(\hat{Q}^{\dag}\hat{c}_{i,0}+\textrm{h.c.})\nonumber \\
&+\sum_{n=0}^{\infty}\omega_{i,n}\hat{c}^{\dag}_{i,n}\hat{c}_{i,n} +(g_{i,n}\hat{c}^{\dag}_{i,n+1}+ \textrm{h.c.)}\bigg].
\end{align}
The definitions of the coefficients in Eq.~\eqref{eq:13} and further details are given in Appendix~\ref{appendix: chain mapping}. These coefficients converge to constants~\cite{10.1063/1.3490188} $\omega_{i,n}\to\omega_{i}=(\omega_{2}-\omega_{1})/2$, $g_{i,n}\to g_{i}=(\omega_{2}-\omega_{1})^{2}/16$ for $n\to\infty$. Here $\ket{\Omega}$ is the particle vacuum $\hat{c}_{1,n}\ket{\Omega} = \hat{c}_{2,n}\ket{\Omega}=0$ for bosons and a particle-hole vacuum $\hat{c}_{1,n}\ket{\Omega} = \hat{c}^{\dag}_{2,n}\ket{\Omega}=0$ for fermions. Note that for the bosonic case, both environments can be described together, giving a single chain geometry as is done in Refs.~\cite{PhysRevLett.123.090402,Le_D__2024}.

\textit{MPS Implementation} - The one-dimensional geometry of the chain mapping is well-suited to MPS techniques where the Hamiltonian is described compactly by a Matrix Product Operator (MPO) and the pure state of the system and environment is described by an MPS. The chain described in Eq.~\eqref{eq:13} is truncated to a finite length $N_{B}$ determined by a Lieb-Robinson bound~\cite{10.1063/1.4940436} which implies that sites further than $\sim g_{i}t$ have a negligible effect on the system dynamics up to time $t$. The time evolution is calculated using the two-site time-dependent variational principle (2TDVP) algorithm~\cite{Fishman2022,Haegeman2016} exploiting charge and magnetization conservation when applicable to enhance efficiency. When mapping multiple chains into a one-dimensional geometry the Hamiltonian will contain terms beyond nearest-neighbour, so it is essential to use a global subspace expansion on the MPS to reduce the projection error~\cite{PhysRevB.102.094315}. This is all implemented using the \texttt{ITensor} package~\cite{Fishman2022}.
% As discussed already, in order to calculate equilibrium correlation functions, the system and environment should reach thermal equilibrium. We do this via real-time equilibration which we optimise by using the state which by definition equilibriates the fastest, $\hat{\rho}_{f} = \hat{\Lambda}^{-1}(\tau_{\rm m}^{\hat{\mathcal{L}}})[\hat{\rho}_{\infty}]$. 

In order for us to describe the combined state of the system and environment as an MPS, the fast initial system state $\hat{\rho}_{f}$ needs to be initialised as a pure state. This is implemented by introducing ancilla and purifying $\hat{\rho}_{f}$ using a Schmidt decomposition 
% \begin{align}
% \hat{\rho}_{f} &= \sum_{i}\lambda_{i}\ket{\psi_{i}}\bra{\psi_{i}}, \\
% \ket{\hat{\rho}_{f}} &= \sum_{i}\sqrt{\lambda_{i}}\ket{\psi_{i}}\ket{\phi_{i}},
% \end{align}
\begin{equation} \label{eq:14}
\hat{\rho}_{f} = \sum_{i}\lambda_{i}\ket{\psi_{i}}_{S}\bra{\psi_{i}}_{S} \to 
\ket{\rho_{f}} = \sum_{i}\sqrt{\lambda_{i}}\ket{\psi_{i}}_{S}\ket{\phi_{i}}_{A},
\end{equation}
such that $\textrm{Tr}_{A}(\ket{\rho_{f}}\bra{\rho_{f}}) = \hat{\rho}_{f}$, where $\ket{\phi_{i}}$ are basis states of the ancilla Hilbert space chosen such that any symmetries of $\hat{\rho}_{f}$ are conserved. For fermions we restrict ourselves to Hamiltonians that conserve total fermion number, meaning $(\hat{N}_{S}\otimes\hat{N}_{A})\ket{\psi_{i}}_{S}\ket{\phi_{i}}_{A} = N\ket{\psi_{i}}_{S}\ket{\phi_{i}}_{A}$, where $N$ is the number of modes in the system and $\hat{N}_{S}$, $\hat{N}_{A}$ are number operators on the system and ancilla modes respectively. The full state is then initialised as $\ket{\psi(0)} = \ket{\rho_{f}}\otimes\ket{\Omega}$ which undergoes unitary evolution of the full Hamiltonian. Once the system is thermalised, we start calculating correlation functions as in Eq.~\eqref{eq:8}. To do this using MPS, we define $C_{A}(t)$ via two separately evolved states 
\begin{equation} \label{eq:16}
\textrm{Tr}(e^{-i\hat{H}t}\hat{A}\rho_{\infty}e^{i\hat{H}t}) = \textrm{Tr}(\ket{\psi^{A}_{\infty}(t)}\bra{\psi_{\infty}(t)}),
\end{equation}
where $\ket{\psi^{A}_{\infty}(t)} = e^{-i\hat{H}t}\hat{A}\ket{\psi_{\infty}}$, $\ket{\psi_{\infty}(t)} = e^{-i\hat{H}t}\ket{\psi_{\infty}}$, $\hat{\rho}_{\infty} = \textrm{Tr}(\ket{\psi_{\infty}}\bra{\psi_{\infty}})$. Once $\tau^{*}_{m}$ is reached, we extrapolate $C_{A}(t>\tau_{m})$ using Eq.~\eqref{eq:10}. For the MPS calculations, we use a time step of $\delta t=0.1/D$ for fermions and $\delta t=0.1/\Delta$ for bosons (see Eq.~\eqref{eq:19} for definition of $\Delta$), a 2TDVP truncation cutoff of $10^{-9}-10^{-10}$, chain lengths of the order $10^2$, generating bond dimensions on the order $10^{2}$. To quantify the error of the extrapolation, we define the error measure $\epsilon(\tau_{m})\equiv \textrm{max}_t|C_{A}(t)-\tilde{C}_{A}(t,\tau_m)|$ as a function of extrapolation time $\tau_{m}$, where $\tilde{C}_{A}(t,\tau_m)$ is the solution obtained via extrapolation as defined in Eq.~\eqref{eq:10} for $t>\tau_m$ and $C_{A}(t)$ is the exact or numerically exact solution. In this sense $\tau_{m}$ acts as a parameter of our methodology, which we increase until convergence, meaning we have reached the true memory timescale $\tau_{m}^{*}$.

\section{Results} \label{sec:Results}
\begin{figure}[t!] % t or b to control placement
  \centering
    \includegraphics[width=1\linewidth]{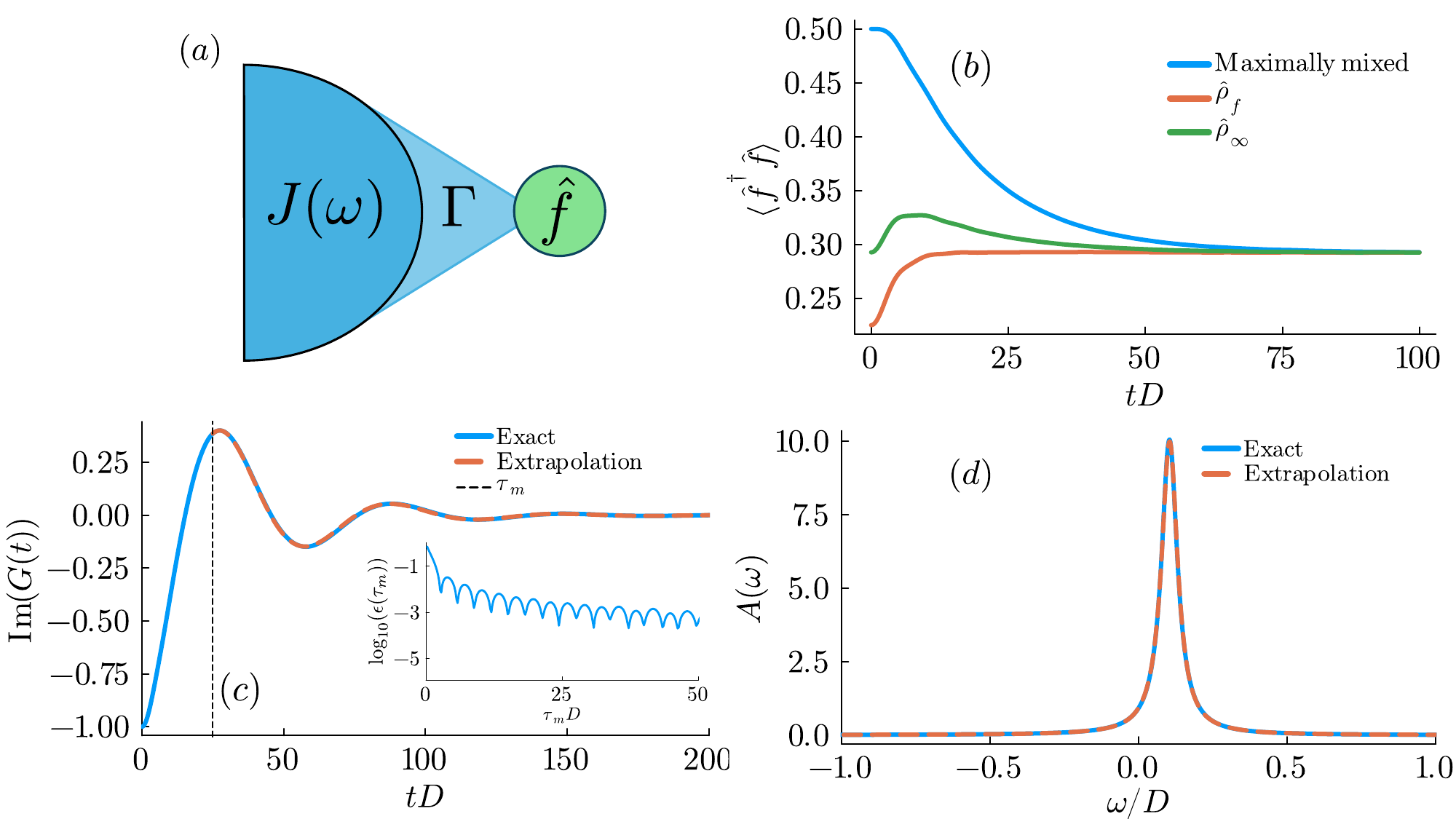}
    \caption{Green's functions and thermalisation for the Resonant Level Model. (a) Schematic of the open system. (b) Thermalisation of the mode density $\langle \hat f^\dagger\hat f\rangle$ towards its steady-state value in $\hat{\rho}_{\infty}$ starting from $\hat{\rho}_{f}$, $\hat{\rho}_{\infty}$ and $\hat{\rho}_\mathbb1$. (c) $G(t)$ calculated exactly and using extrapolation from $t=25/D$, with the inset showing the maximum error for the extrapolation as a function of $\tau_{m}$. (d) Comparison of $A(\omega)$ calculated from the extrapolated $G(t)$ with the exact solution. Parameters: $\Gamma=0.05D,\beta=10/D,\mu=0,\epsilon=0.1D$, maximum bond dimension $\chi_{\textrm{max}}=300$, $N_{B} = 210$. }
  \label{fig:RLM plots}
\end{figure}

We begin by considering spinless fermions coupled to a fermionic reservoir. In the non-interacting limit this system can be solved exactly, allowing us to benchmark our method. For simplicity, we assume a semi-elliptical spectral function
\begin{align} \label{eq:17}
    \mathcal{J}(\omega) &= \frac{2\Gamma}{\pi^2}\sqrt{1-(\omega/D)^2},
\end{align}
such that $\omega_1,\omega_2 = -D,D$. Taking the system as being described by mode operators $\hat{f}_i$, our focus is on computing the retarded single-particle Green's function $G_{i}(t) = -i\textrm{Tr}(\{\hat{f}_{i}(t),\hat{f}_{i}^{\dag}\}\hat{\rho}_{\infty})$ and its corresponding spectral function defined via $A_{i}(\omega) \equiv -\frac{1}{\pi}\textrm{Im}\int_{0}^{\infty}dt~e^{i\omega t}G_{i}(t)$. 

\textit{Resonant Level Model} - We first consider a system comprising of a single fermion mode with $\hat{H}_{S}=\epsilon\hat{f}^{\dag}\hat{f}$, coupled to the environment via $\hat{Q}=\hat{f}$ as depicted in Fig.~\ref{fig:RLM plots}(a). This entirely non-interacting setup is an ideal test case since it has an efficient numerically exact solution based on the single-particle correlation matrix to compare against, as outlined in Appendix~\ref{appendix: non interacting fermions}, yet it remains a non-trivial calculation for MPS techniques with a complexity similar to later interacting examples. Step 1 of our approach involves thermalisation $\hat{\rho}(0) \to \hat{\rho}_{\infty}$. Generically this relaxation can take a long time depending on the initial state. Here we minimise this by using $\hat{\rho}_{f}$. This speed up is illustrated in Fig.~\ref{fig:RLM plots}(b), which compares the relaxation dynamics of the density for three different initial states: the fast state $\hat{\rho}_{f}$, the steady-state itself $\hat{\rho}_{\infty}$ and the maximally mixed state $\hat{\rho}_\mathbb{1}$. The fast state thermalises within $t \sim 25/D$, while $\hat{\rho}_{\infty}$ takes $t \sim 50/D$ to relax, which is an example of the extreme NMQMpE~\cite{PhysRevLett.134.220403}, and the maximally mixed state takes significantly longer, relaxing only by $t \sim 100/D$. This demonstrates how the required time, and therefore computational resources, for thermalisation can be minimised by leveraging the NMQMpE, displaying it's practical significance in addition to its fundamental scientific interest. 

\begin{figure}[t!] % t or b to control placement
  \centering
    \includegraphics[width=1\linewidth]{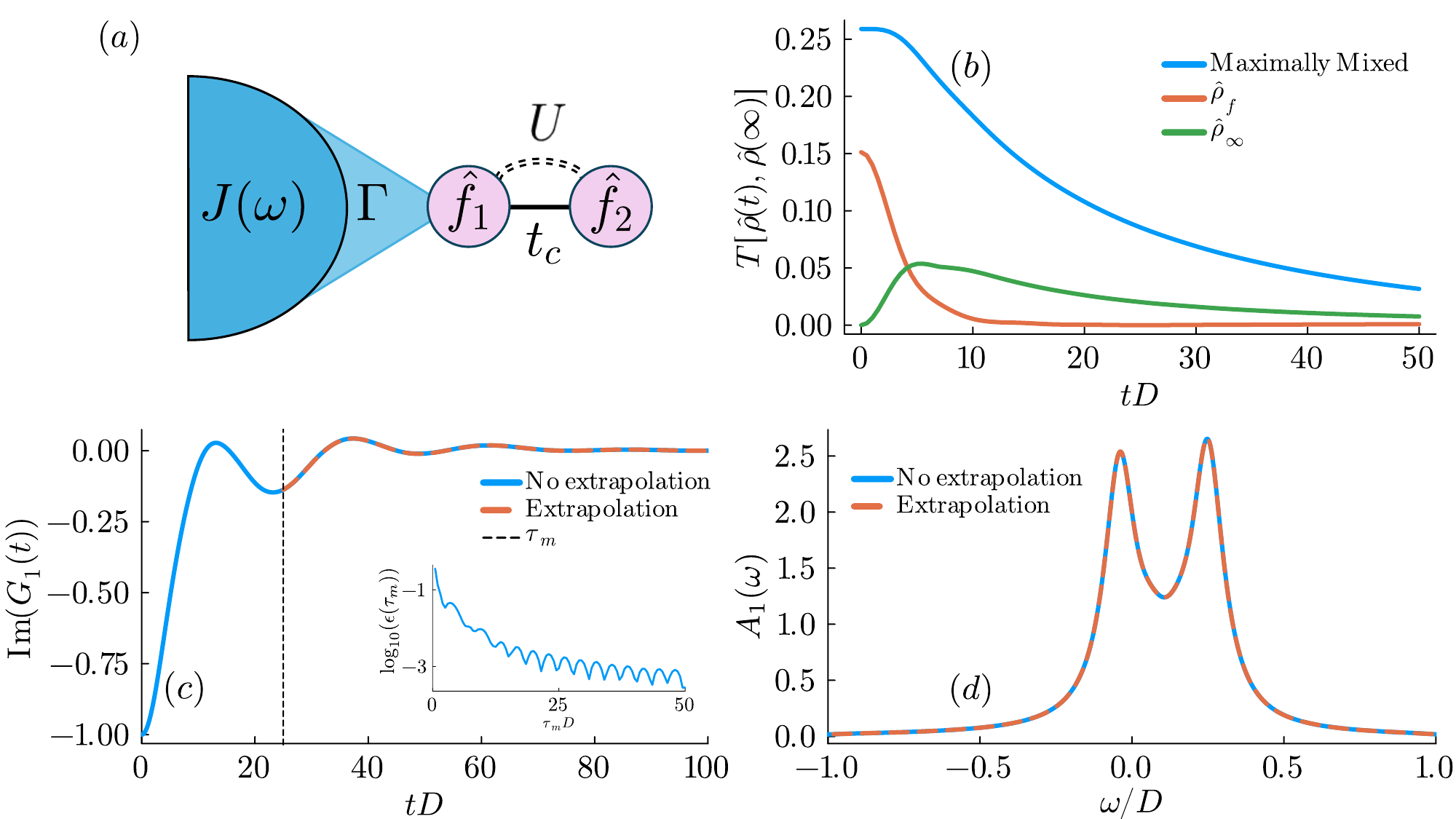}
  \caption{Green's functions and thermalisation for two interacting fermionic modes. (a) Schematic of the two-mode open system. (b) Thermalisation in terms of trace distance starting from $\hat{\rho}_{f}$, $\hat{\rho}_{\infty}$ and the maximally mixed state. (c) $G_{1}(t)$ calculated using the direct MPS method versus extrapolation from $t=25/D$, with the inset showing the maximum error of the extrapolation as a function of $\tau_{m}$. (d) Comparison of $A_1(\omega)$ calculated from the extrapolated $G_1(t)$ with the near-exact direct MPS result. Parameters: $\Gamma = 0.1D, \beta = 10/D,\mu = 0.1D,t_c = 0.05D, U = 0.2D$, maximum bond dimension $\chi_{\textrm{max}}=300$, $N_{B} = 150$.}
  \label{fig:Two coupled modes plots}
\end{figure}

\begin{figure*}[t!] % t or b to control placement
  \centering
    \includegraphics[width=1\linewidth]{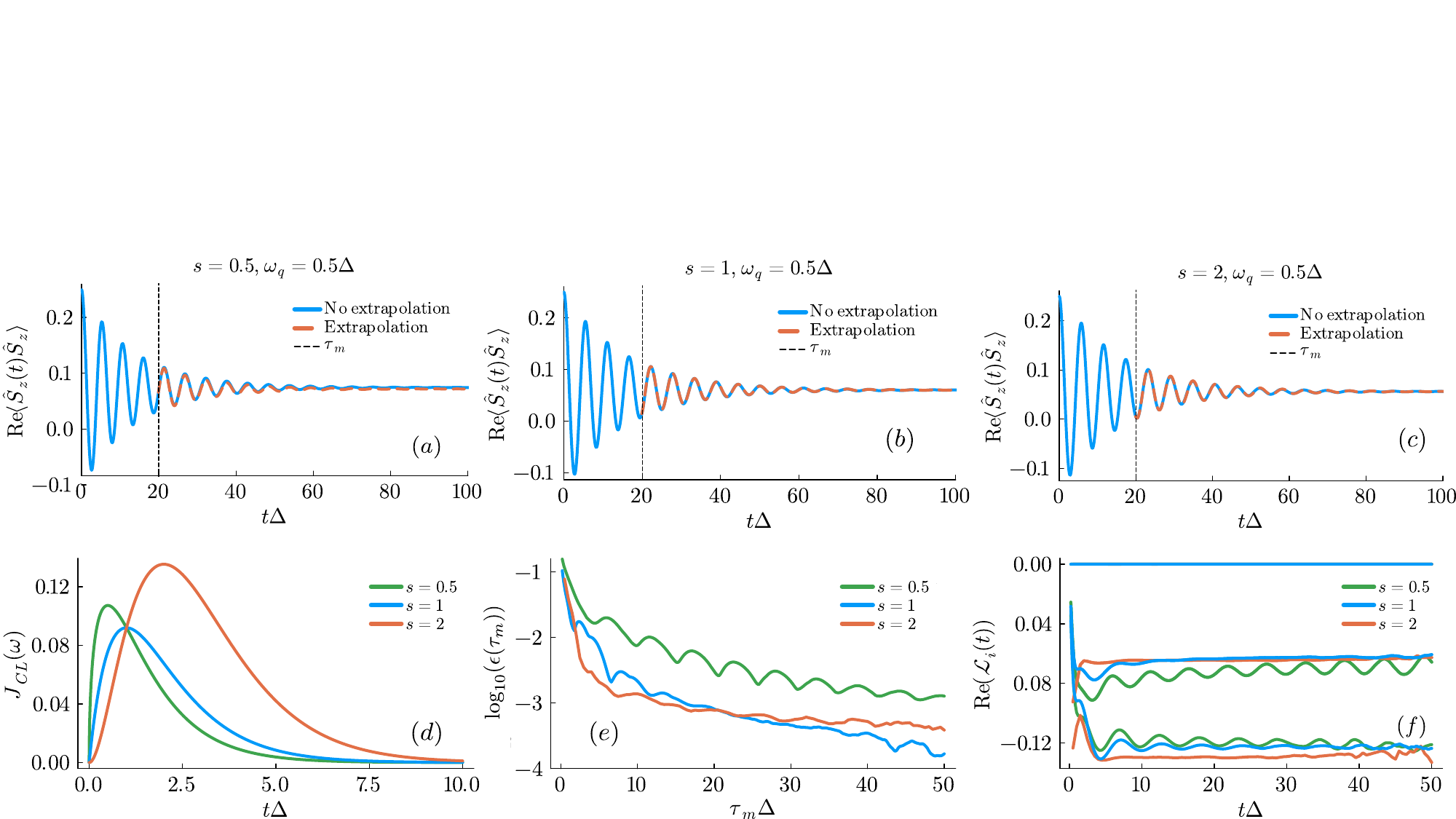}
    \caption{$\langle \hat{S}_{z}(t)\hat{S}_{z}\rangle$ for the spin-boson model and its thermalisation for sub-ohmic, ohmic and super-ohmic spectral densities. (a)-(c) $\langle \hat{S}_{z}(t)\hat{S}_{z}\rangle$ with ohmicity $s=0.5,1,2$ respectively, calculated using direct MPS (blue) and extrapolation from $t=20/\Delta$ (orange). (d) The three spectral densities used. (e) Errors for the three cases in terms of the maximum deviation between the direct MPS result and the extrapolated result. (f) Shows the real part the spectrum of $\hat{\mathcal{L}}(t)$ for the three cases. Parameters: $\Gamma=0.25\Delta,\beta=\infty,\omega_{q}=0.5\Delta$, $N_B=800$, maximum bond dimension $\chi_{\textrm{max}}=50,20,15$ for $s=0.5,1,2$ respectively.}
  \label{fig:Spin Boson plots}
\end{figure*}

In Fig.~\ref{fig:RLM plots}(c) the Green’s function $G(t) = -i\textrm{Tr}(\{\hat{f}(t),\hat{f}^{\dag}\}\hat{\rho}_{\infty})$ of the system mode is shown. This was calculated using two approaches: the exact single-particle solution and MPS-based evolution which computes $G(t)$ up to $t = \tau_m = 25/D$, after which our extrapolation scheme is applied for all later times. The two methods show excellent agreement, establishing the accuracy of the MPS calculation up to $t=\tau_m$ comprising step 2, and the validity of our extrapolation beyond this, making up step 3. The power of this method is that we are able to extract the full evolution of $G(t)$ for arbitrarily long times after $t=\tau_{m}$ without the need for costly MPS calculations. Instead the calculation is reduced to trivial $2\times 2$ matrix operations \footnote{For a general system of $N_s$ modes, the size of the propagator $\mathcal{L}_{m}$ is $d^{2N_{s}}\times d^{2N_{s}}$ where $d$ is the size of the dimension of the each mode, but for the resonant level model the Hamiltonian is number conserving meaning $\hat{\rho}(t)$ is diagonal in the number basis, reducing the state space by a factor of 2.}. By removing the need to track the full many-body state of the system and environment for $t > \tau_m$ we avoid the well known `entanglement barrier'~\cite{PhysRevLett.124.137701} typical in many-body quantum dynamics, allowing long timescales to be reached. Indeed, for this example, and the cases shown later in Figs.~\ref{fig:Two coupled modes plots},~\ref{fig:Spin Boson plots}, we find an order of magnitude difference in the wall-clock time taken for completing both the thermalisation step 1 and the Green's function calculation in step 2 using our methodology compared to a direct MPS calculation that doesn't exploit either the NMQMpE or the extrapolation.

The error measure $\epsilon(\tau_{m})$ is shown in the inset of Fig.~\ref{fig:RLM plots}(c) where an error of $\sim 10^{-3}$ is reached by $\tau_{m}\sim 25/D$. In frequency space $G(\omega)$, shown in Fig.~\ref{fig:RLM plots}(d), displays a peak centred at the energy of the impurity with the spread induced by the environment coupling. As expected, we see excellent agreement with the exact solution.

\textit{Two interacting modes} - 
We now consider an interacting setup with no exact solution. Specifically, two coupled fermionic modes with a density-density interaction described by a Hamiltonian $\hat{H}_{S} = t_{c}(\hat{f}_{2}^{\dag}\hat{f}_{1}+\hat{f}_{1}^{\dag}\hat{f}_{2})+U\hat{f}_{1}^{\dag}\hat{f}_{1}\hat{f}_{2}^{\dag}\hat{f}_{2}$, coupled to the environment via $\hat{Q}=\hat{f}_{1}$ as depicted in Fig.~\ref{fig:Two coupled modes plots}(a). Isolated this system has single-particle eigenstates $\ket{\pm} = \tfrac{1}{\sqrt{2}}(\ket{1,0} \pm \ket{0,1})$ with energies $\epsilon = \pm t_c$, an empty eigenstate $\ket{0,0}$ at energy $\epsilon = 0$, and a fully occupied eigenstate $\ket{1,1}$ at energy $\epsilon = U$, where $\ket{n_1,n_2} = (\hat f_1^\dagger)^{n_1}(\hat f_2^\dagger)^{n_2}\ket{\rm vac}$ denotes system Fock states. 
Here we rely on the exclusive direct application of MPS for benchmarking and validation of our acceleration scheme. In Fig.~\ref{fig:Two coupled modes plots}(b) the relaxation is shown for the same initial states, $\hat{\rho}_{f}$, $\hat{\rho}_{\infty}$ and $\hat{\rho}_\mathbb{1}$, expressed in terms of the trace distance $T[\hat{\rho}(t),\hat{\rho}(\infty)]\equiv \frac{1}{2}\textrm{Tr}[\sqrt{(\hat{\rho}(t)-\hat{\rho}(\infty))^{\dag}(\hat{\rho}(t)-\hat{\rho}(\infty))}]$. The NMQMpE again emerges, with the fast state thermalising by $t<20/D$ compared to the maximally mixed state and $\hat{\rho}_{\infty}$ that do not reach thermalisation by the final $t=50/D$ displayed. In Fig.~\ref{fig:Two coupled modes plots}(c) the Green's function $G_{1}(t)$ for the first system mode is reported. Importantly, the presence of interactions does not affect the accuracy of our approach: the extrapolated $G_{1}(t)$ matches the near-exact direct MPS result extremely well, with the inset showing a maximum error of $\sim10^{-3}$ by $\tau_{m}/D\sim 25$. Correspondingly the spectral features in $A_1(\omega)$ for this system are accurately reproduced in Fig.~\ref{fig:Two coupled modes plots}(d), which are dominated by two peaks arising from transitions from the groundstate $\ket{-}$ of $\hat{H}_S$ up to $\ket{1,1}$ at $\omega_- \approx U + t_c$, and from the excited state $\ket{+}$ down to $\ket{0,0}$ at $\omega_- \approx -t_c$.

\begin{figure}[t!] % t or b to control placement
  \centering
    \includegraphics[width=1\linewidth]{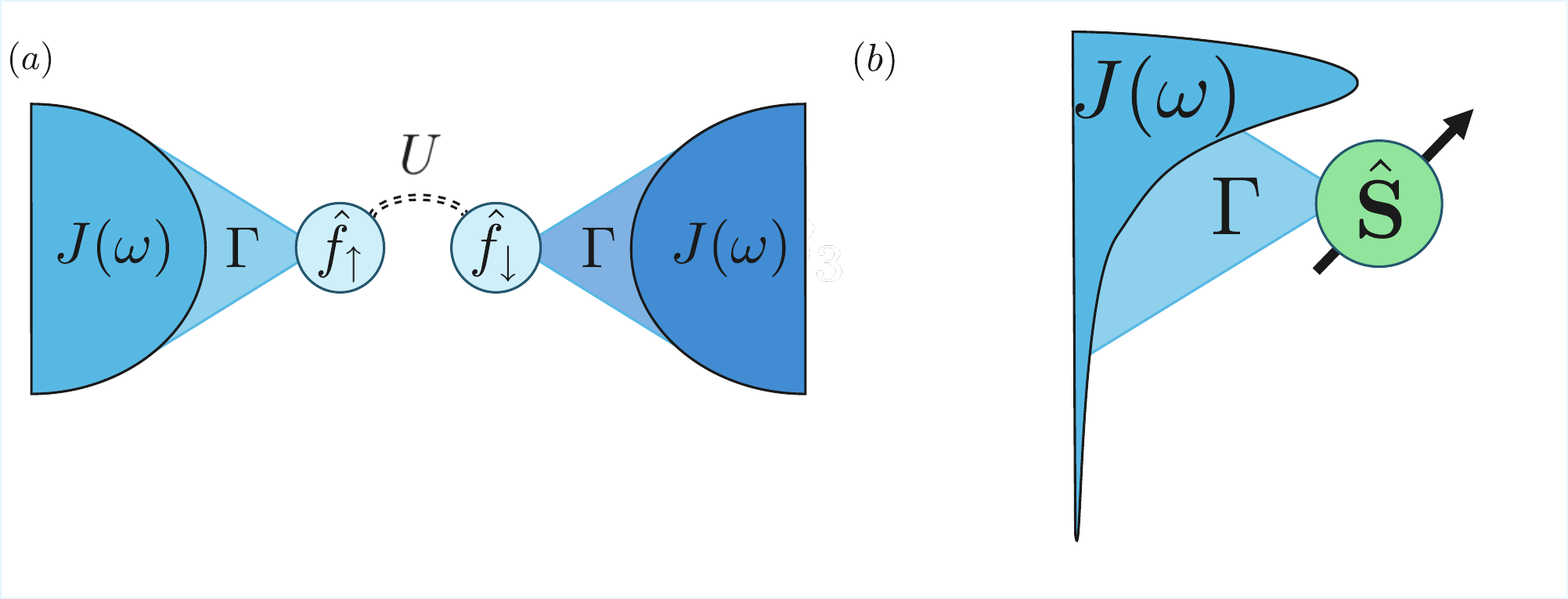}
  \caption{Schematics for (a) the single impurity Anderson model and (b) the spin-boson model.}
  \label{fig:Schematics for SIAM and spin boson}
\end{figure}

\textit{Anderson impurity model} - We now consider a more challenging paradigmatic example, the particle-hole symmetric single impurity Anderson model (SIAM), with Hamiltonian
\begin{equation} \label{eq:18}
\hat{H}_{S} = U\hat{f}_{\uparrow}^{\dag}\hat{f}_{\uparrow}\hat{f}_{\downarrow}^{\dag}\hat{f}_{\downarrow} - \frac{U}{2}(\hat{f}_{\uparrow}^{\dag}\hat{f}_{\uparrow}+\hat{f}_{\downarrow}^{\dag}\hat{f}_{\downarrow}),
\end{equation}
as depicted in Fig.~\ref{fig:Schematics for SIAM and spin boson}(a). For strong interactions $U$ and low temperatures the model exhibits complex memory effects arising from the Kondo effect, where the impurity entangles with the environment electrons screening its local moment~\cite{PhysRevB.91.085127,PhysRevB.88.094306,Coleman_2015}. This results in a peak at the Fermi energy in the spectral function $A_{\sigma}(\omega)$ for $\sigma =\, \uparrow,\downarrow$. The environment electrons can be understood as two separate environments of the form Eq.~\eqref{eq:12} for each spin component $\sigma$, each coupled to the associated spin of the system electron, such that the spin up and spin down electrons only interact through the system impurity, with $\hat{Q}_{\uparrow}=\hat{f}_{\uparrow}$, $\hat{Q}_{\downarrow}=\hat{f}_{\downarrow}$. We again use a semi-elliptical spectral function for both spin components and use the same parameters as in Ref.~\cite{PhysRevB.104.014303} with a strong coupling and interaction $\Gamma=\pi D/20,\; U=5\Gamma$ with the Fermi energy $\mu=0$~\footnote{Note that the definition of $\Gamma$ used in this works differs from the definition in Ref.~\cite{PhysRevB.104.014303}.}. We consider two cases, $\beta=2.5/D$ and $\beta=20/D$ which show the emergence of the Kondo peak.

\begin{figure}[t!] % t or b to control placement
  \centering
    \includegraphics[width=1\linewidth]{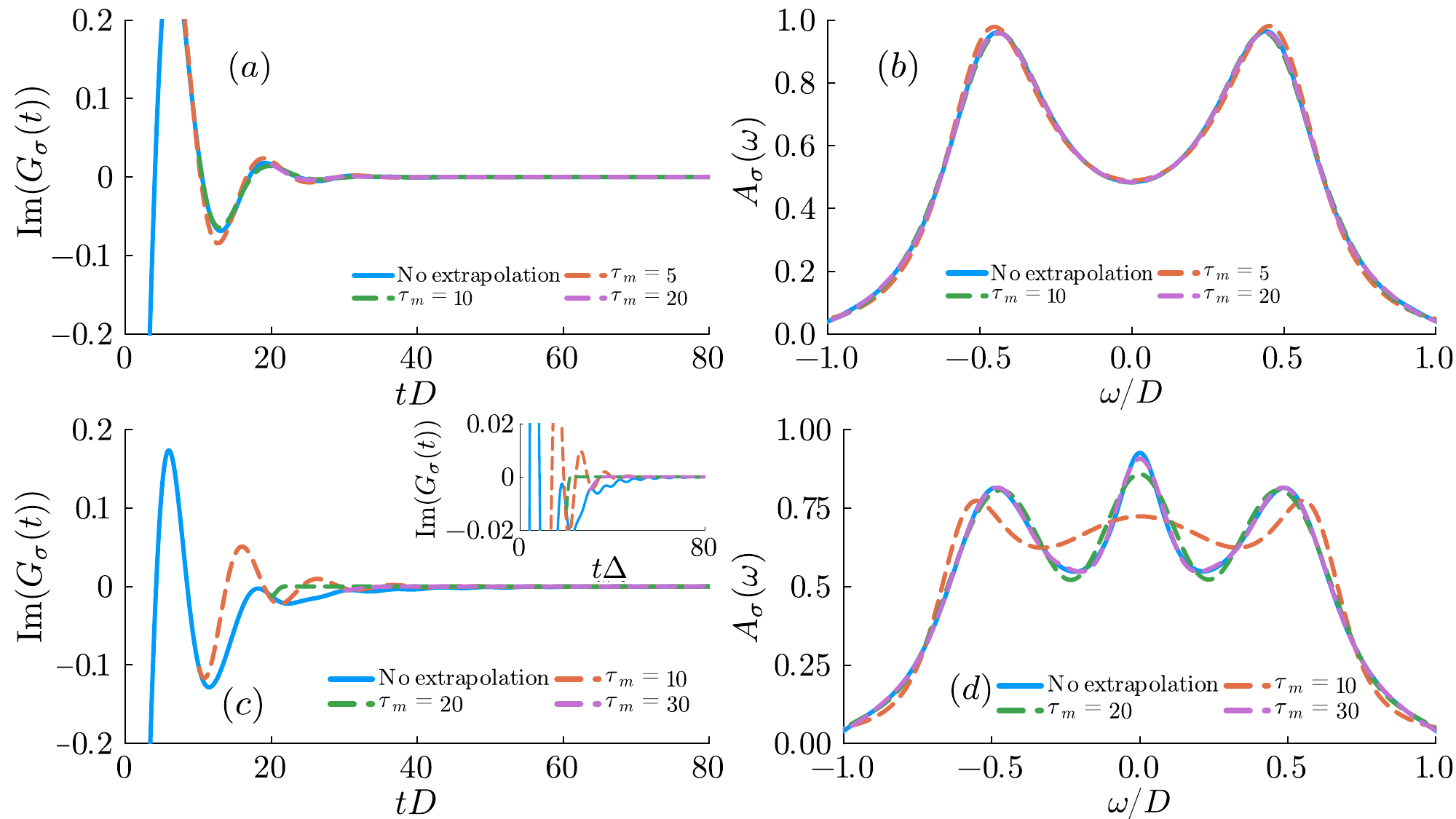}
    \caption{Green's functions and spectral functions for the SIAM. (a) $G_\sigma(t)$ for $\beta=2.5/D$, calculated using direct MPS versus Eq.~\eqref{eq:10} for various choices of $\tau_{m}$. (b) The associated spectral function. (c) $G_\sigma(t)$ for $\beta=2.5/D$, calculated using direct MPS versus Eq.~\eqref{eq:10} for various choices of $\tau_{m}$.  (d) The associated spectral function $A_\sigma(\omega)$. Parameters: $\Gamma=\pi D/20,\;U=5\Gamma,\;\mu=0$, $N_{B}=80$, maximum bond dimension $\chi_{\textrm{max}}=150$ (same as Ref.~\cite{PhysRevB.104.014303}).}
  \label{fig:SIAM plots}
\end{figure}

For both $\beta$'s, we find little difference in thermalisation times of $\hat{\rho}_{f}$ and $\hat{\rho}_{\infty}$, meaning the extreme NMQMpE effect has become negligible. However, both these states are only available apriori for step 1 due to computing the dynamical map, and still provide significant speed ups compared to using a generic initial state which can display very slow relaxation, as shown in Appendix~\ref{appendix:SIAM thermalisation}. In Fig.~\ref{fig:SIAM plots}(a) the Green's function $G_\sigma(t)$ for either $\sigma$ is reported for $\beta=2.5/D$ computed directly using MPS and compared with using Eq.~\eqref{eq:10} for various extrapolation times. Using $\tau_{m} = 5/D$ roughly captures the dynamics, with the larger $\tau_{m}$ choices rapidly converging to the true solution. This is reflected in frequency space as shown in Fig.~\ref{fig:SIAM plots}(b), where both $\tau_{m}=10/D$ and $\tau_{m}=20/D$ capture the spectral function well, showing two Hubbard sidebands symmetrically located at $\omega \approx\pm (\tfrac12 + \tfrac{\Gamma}{\pi D})U$. However, this case demonstrates when our extrapolation scheme is of limited advantage due to a lack of separation of timescales. Specifically, the timescale over which Eq.~\eqref{eq:10} is valid is similar to the timescale over which the Green's function itself has almost decayed away, meaning there is little meaningful variation in $G_\sigma(t)$ to extrapolate.

In Figs.~\ref{fig:SIAM plots}(c),(d) the associated results for the $\beta=20/D$ case are shown. In this case the Kondo peak has emerged at the Fermi energy $\omega=0$. Here we can see Eq.~\eqref{eq:10} fails to capture the dynamics of $G_\sigma(t)$ for all chosen extrapolation times, as is clear in the inset zoom in of $G_\sigma(t)$. For example, using $\tau_m = 20/D$ we see $G_\sigma(t)$ display an immediate discontinuity in its derivative at the extrapolation point $t=\tau_m$. If $\hat{\mathcal{L}}(\tau_m)$ was exclusively governing the dynamics of $\hat{\rho}(t)$ at $t=\tau_m$ then the extrapolation would be smooth. In this case we therefore have $\mathcal{I}(\tau_m)\neq0$ so $\tau_m < \tau_{m}^{I}$, and thus $\tau_{m}^{*}$ has not been reached. The initial system-environment correlations of the equilibrium state have a persistent relevance reflecting the long-lasting memory effect associated with the emergence of Kondo physics. The failure of the extrapolation is less obvious in frequency space which appears to show convergence to the true solution, but this is only due to the accurate direct MPS solution capturing more of the true time-dependence and decreasing contribution of the extrapolation with increasing $\tau_{m}$. Consequently, for the SIAM step 1 of our acceleration scheme still provides a speed-up, but the advantages of step 3 are diminished.

\textit{Spin-boson model} - For our final example we switch to the spin-boson model, depicted in Fig.~\ref{fig:Schematics for SIAM and spin boson}(b), another paradigmatic model of open quantum systems frequently used to study non-Markovian dynamics and benchmark methods. In this case the system is a single spin-1/2 degree of freedom governed by a Hamiltonian 
\begin{equation} \label{eq:19}
\hat{H}_{S} = \omega_{q}\hat S_{z}+\Delta\hat S_{x},
\end{equation}
where $\omega_{q}$ defines the bias between the two spin states, $\Delta$ is the tunneling strength and $\hat{\bf S} = (\hat{S}_{x},\hat{S}_y,\hat{S}_z)$ are spin-1/2 operators. The system is coupled to a bosonic environment via $\hat{Q} = \hat{S}_z$. We restrict ourselves to the zero temperature regime to focus on non-Markovian dynamics, which is equivalent to setting the initial state of the bosonic environment to be in the vacuum \footnote{Technically, thermal states are ill-defined for sub-ohmic environments. The reason for this is the quantity $n(\omega)J(\omega)$ diverges at $\omega\to0$ where $n(\omega)$ is the Bose occupation factor and $J(\omega)$ is the spectral function. This quantity is the environment correlation function~\cite{PhysRevA.93.062114} which partially defines the thermal environment. This is not an issue if we take an empty environment, which is formally the zero-temperature limit of a sub-ohmic environment that has been appropriately regularised near $\omega=0$ to have a well defined correlation function.}. Despite its apparent simplicity, this model has no analytical solution for $\Delta\neq0$. We consider the Caldeira and Leggett model to describe the spectral density of the bosonic environment, given by
\begin{equation} \label{eq:20}
J_{CL}(\omega) = \Gamma\omega_{c}^{1-s}\omega^{s}e^{-\omega/\omega_{c}},
\end{equation}
where $\Gamma$ characterises the system-environment coupling strength and $\omega_{c}$ is the cutoff frequency. Since there is formally no finite bandwidth for this spectral density we use $\Delta=1$ to define our energy scale for this model. The parameter $s$ in $J_{CL}(\omega)$ characterises the nature of the environment as depicted in Fig.~\ref{fig:Spin Boson plots}(d), with $0<s<1$ being sub-ohmic, $s=1$ ohmic, and $s>1$ super-ohmic. The sub-ohmic case is a particularly challenging regime due to its higher density of low-frequency excitations resulting in a longer memory time~\cite{10.1063/5.0235741}, particularly at zero temperature. 

In this model we focus on the two-time correlation $\langle \hat{S}_{z}(t)\hat{S}_{z}\rangle$. In Fig.~\ref{fig:Spin Boson plots}(a)-(c) $\textrm{Re}(\langle \hat{S}_{z}(t)\hat{S}_{z}\rangle)$ is reported, with $\tau_{m}=20/\Delta$ used for the extrapolation in all three cases. As can been seen there is excellent agreement between the extrapolation and direct MPS results for the ohmic and super-ohmic cases, but some deviation observed for the sub-ohmic case. This is quantified in Fig.~\ref{fig:Spin Boson plots}(e) where the maximum error for ohmic and super-ohmic is $\sim10^{-3}$ within $\tau_{m}=20/\Delta$. The sub-ohmic case fails to reach this level of accuracy for any $\tau_m \leq 50/\Delta$ reflecting the difficulty of this regime. 

The differing behaviour of the extrapolation with $s$ is illustrated more starkly by examining the real part of the spectrum of $\hat{\mathcal{L}}(t)$ which characterises the time dependent damping rates of the system dynamics. This is reported in Fig.~\ref{fig:Spin Boson plots}(f) where we see the fastest convergence for the super-ohmic environment within $t\sim 10/\Delta$, while the ohmic environment converging by approximately $t\sim 20/\Delta$. In contrast the spectrum for the sub-ohmic case shows persistent oscillations resulting in persistent deviations when applying the extrapolation scheme. The deviations from stationarity at large times for the faster decay mode for all three cases are due to this mode being projected out in $\hat{\Lambda}(t)$ and thus their decay rate becomes undefined.

\begin{figure}[t!] % t or b to control placement
  \centering
    \includegraphics[width=1\linewidth]{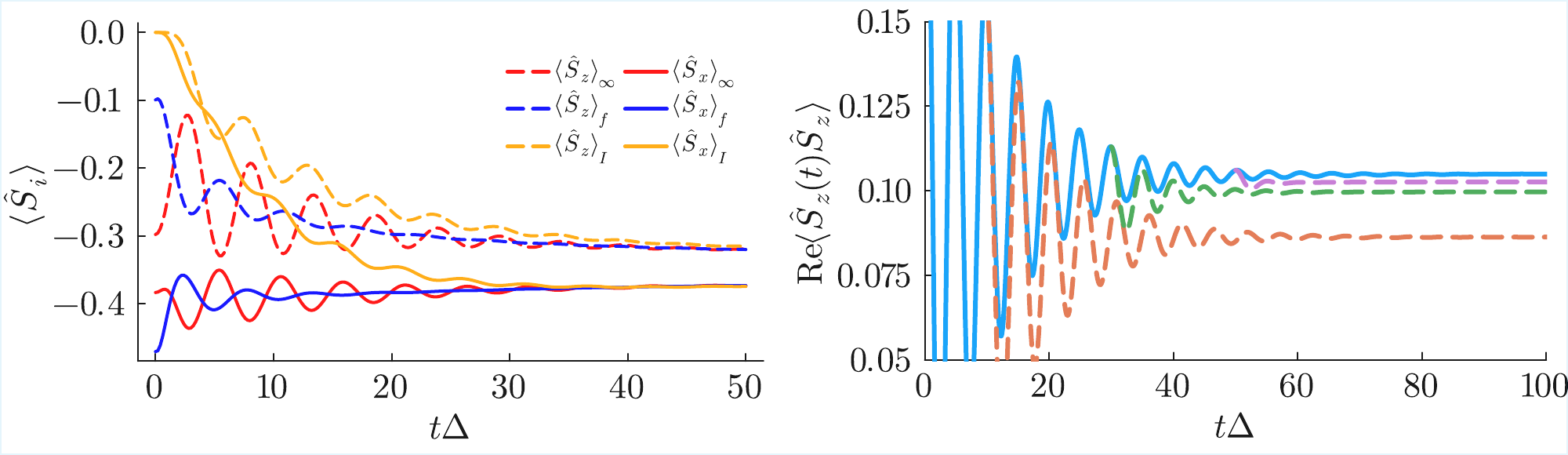}
    \caption{(a) The thermalisation of the spin components $\langle \hat{S}_{i}(t)\rangle$ for the spin boson model deep in the sub-ohmic regime, with $s=0.1$, for the three initial initial states $\hat\rho_f$, $\hat\rho_\infty$ and $\hat\rho_\mathbb1$. (b) The extrapolation of $\langle \hat{S}_{z}(t)\hat{S}_{z}\rangle$ with increasing $\tau_m$. Parameters: $\Gamma=0.25\Delta,\beta=\infty,\omega_{q}=0.5\Delta$, $N_{B}=800$, maximum bond dimension $\chi_{\textrm{max}}=120$.}
  \label{fig:Sub Ohmic plots}
\end{figure}

\textit{Strongly sub-ohmic} - For the three cases considered for the spin-boson model, our methodology works well. We now consider a deeply sub-ohmic case where the dynamics are strongly non-Markovian. For this we consider the same parameters as before, but with $s=0.1$. In Fig.~\ref{fig:Sub Ohmic plots}(a) the thermalisation process for each spin component is shown for $\hat{\rho}_{f}$ calculated from $\tau_{m}=20/\Delta$, $\hat{\rho}_{\infty}$ and $\hat{\rho}_\mathbb1$ initial states. All three states relax on roughly the same timescale, meaning there is no separation between the relaxation time and the memory times here, resulting in no discernible extreme NMQMpE to take advantage of. This is further demonstrated when applying Eq.~\eqref{eq:10} to the calculation of $\langle \hat{S}_{z}(t)\hat{S}_{z}\rangle$. Each choice of $\tau_{m}$ immediately deviates from the true correlation function once extrapolation is applied causing them to reach a different steady state value. The deviations shrinks with increasing $\tau_{m}$, but reflects the lack of convergence of $\hat{\mathcal{L}}(t)$ on the timescales examined, indicating very strong memory effects. 

\section{Conclusion} \label{sec:Conclusion}
In this work we have extended the dynamical map framework~\cite{strachan2024extracting}, previously developed for single-time observables, to the calculation of equilibrium two-time correlation functions in non-Markovian open quantum systems. By combining the central idea of extrapolation from Ref.~\cite{strachan2024extracting} with the non-Markovian quantum Mpemba effect~\cite{PhysRevLett.134.220403}, we have introduced a numerically efficient scheme that minimizes computational costs and enables the calculation of two-time dynamics from relatively short-time data.

We benchmarked this approach across a variety of paradigmatic models for both fermionic and bosonic environments. For the spinless fermionic models considered, both with and without interactions, the extrapolation reproduces the Green's function with high accuracy, significantly reducing simulation resources, where we found order of magnitude speed ups. The method succeeds in the SIAM at moderate temperatures, though the advantages of the extrapolation diminish in regimes lacking a clear separation of timescales between relaxation and memory owing to the emergence of the Kondo effect. For the spin-boson model, our approach accurately describes ohmic, super-ohmic and moderately sub-ohmic environments, but struggles in the strongly sub-ohmic regime where memory effects persist indefinitely.

Taken together, these results establish the use of dynamical maps as a powerful tool for computing equilibrium two-time correlation functions in non-Markovian settings. While the acceleration proposed is not universally applicable, the approach can provide substantial computational gains whenever environmental memory saturates on timescales shorter than relaxation. Future work will focus on developing this approach to more accurately describe cases with persistent memory, for example adding additional modes at zero energy that fully capture the Kondo cloud formation, such that the effect of the remaining environment on the effective system becomes more Markovian. This could be very beneficial for impurity solvers in DMFT~\cite{RevModPhys.78.865,RevModPhys.68.13,Vollhardt2012} where the use of dynamical maps for the impurity is still feasible since the size of the impurities typically do not exceed $\sim 14$ fermionic modes. 

\begin{acknowledgments}
    S.R.C. gratefully acknowledges financial support from UK's Engineering and Physical Sciences Research Council (EPSRC) under grant EP/T028424/1. A.P acknowledges funding from Seed Grant from IIT Hyderabad, Project No. SG/IITH/F331/2023-24/SG-169. A.P also acknowledges funding from Japan International Coorperation Agency (JICA) Friendship 2.0 Research Grant, and from Finnish Indian Consortia for Research and Education (FICORE).
\end{acknowledgments}
\clearpage

\appendix 
\onecolumngrid

\section{Extracting the Dynamical Map} \label{appendix:Extracting map}

\begin{figure}[h!]
  \centering
  \includegraphics[width=0.8\textwidth]{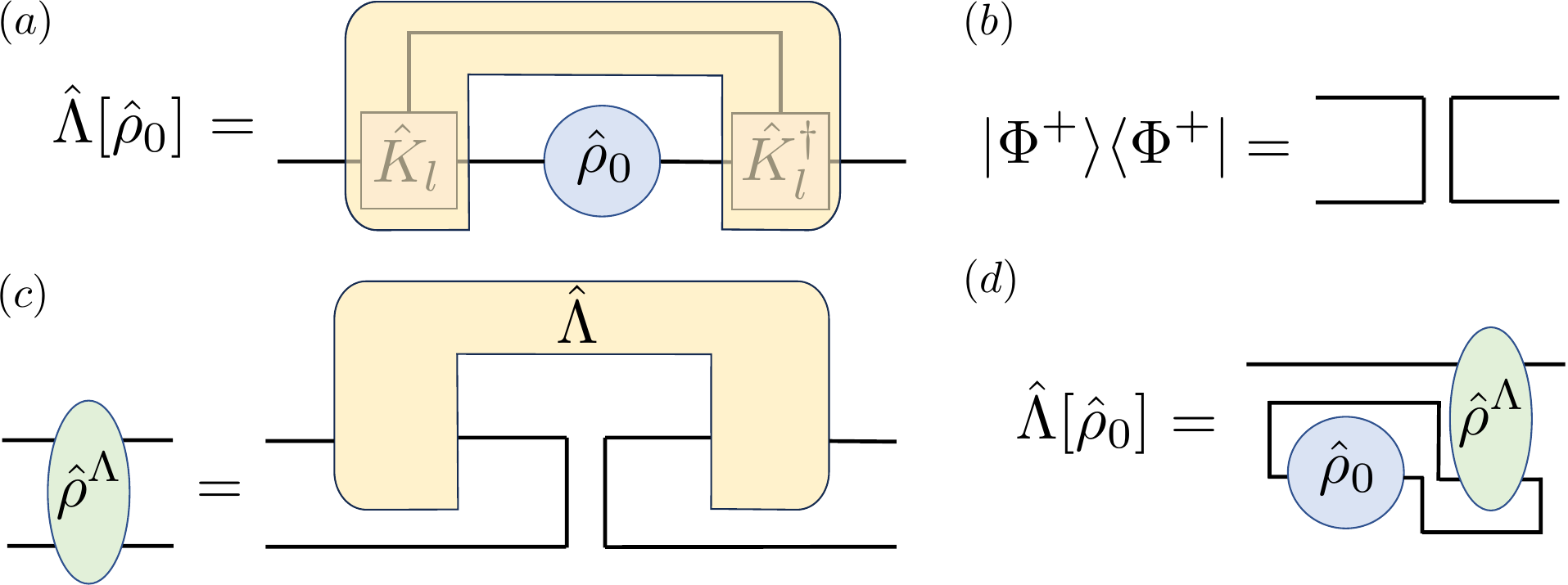}
  \caption{Tensor network type diagrams of the Choi-Jamiolkowski isomorphism. (a) The application of a CPTP map $\hat{\Lambda}[\hat{\rho}_0]$ can be viewed as a cap shaped tensor encompassing $\hat{\rho}$. Internally the cap tensor could be decomposed into a conjugation by Kraus operators $\hat{K}_l$ as shown. (b) Owing to its perfect correlations the maximally entangled state $\ket{\Phi^+}$ from Eq.~\eqref{eq:A1} is a bent line. (c) The Choi-Jamiolkowski isomorphism follows from inputting $\ket{\Phi^+}$ into $\hat{\Lambda}$, which diagrammatically corresponds to bending the input legs of $\hat{\Lambda}$ outwards to form $\hat{\rho}^\Lambda$ as in Eq.~\eqref{eq:A2}. (d) The application of the map $\hat{\Lambda}[\hat{\rho}_0]$ is extracted from $\hat{\rho}^\Lambda$ by contracting the appropriate legs consistent with (a), and inserting a factor of the dimension $d$ to get the graphical version of Eq.~\eqref{eq:A3}.}
  \label{fig:isomorphism}
\end{figure}

Consider a $d$-level system each with basis states $\ket{s}$ for which the application of any completely positive trace preserving map (CPTP) map $\hat{\Lambda}$ to an initial state $\hat{\rho}_{0}$ can be drawn as in Fig.~\ref{fig:isomorphism}(a). The trick of the isomorphism is to add auxilliary replica degrees of freedom $A_{S}$ for the system $S$, and initialize the system and its auxiliary replica in a perfectly correlated maximally entangled state
\begin{align}
   \ket{\Phi^{+}} = \frac{1}{\sqrt{d}}\sum_{s=1}^{d}\ket{s}_{S}\otimes\ket{s}_{A_{S}}, \label{eq:A1}
\end{align}
which has a diagrammatic form shown in Fig.~\ref{fig:isomorphism}(b). Since $\hat{\Lambda}$ is completely positive applying it to the system alone is guaranteed to generate a valid density operator in the space of system + replica operators~\footnote{Conversely, for any nonnegative operator we can also to a corresponding quantum map.} given by
\begin{align}
    \hat{\rho}^{\Lambda} = (\hat{\Lambda} \otimes \mathds{1})\big[\ket{\Phi^{+}}\bra{\Phi^{+}}\big], \label{eq:A2}
\end{align}
as depicted in Fig.~\ref{fig:isomorphism}(c). Armed with $\hat{\rho}^{\Lambda}$ the action of the map $\hat{\Lambda}$ then follows
\begin{equation}
    \hat{\Lambda}[\hat{\rho_{0}}] = d\;\textrm{Tr}_{A_S}(\mathbb{1}\otimes\hat{\rho}_{0}^{\textrm{T}}\hat{\rho}^{\Lambda}), \label{eq:A3}
\end{equation}
as summarized in Fig.~\ref{fig:isomorphism}(d). The isomorphism straightforwardly extends to maps applied to $N$ system sites by using the state
\begin{equation} \label{eq:A4}
   \ket{\Phi^{+}_N} = \frac{1}{\sqrt{d^N}}\otimes_{j=1}^{N}\bigg(\sum_{s_j=1}^{d}\ket{s_j}_{S}\otimes\ket{s_j}_{A_S}\bigg), 
\end{equation}
where each site is maximally entangled with its own auxiliary replica. The application of the isomorphism to many-body fermionic systems also follows analogously once some subtle details are taken care of \cite{strachan2024extracting}.

\section{Chain mapping} \label{appendix: chain mapping}

Here we provide more details of the chain mapping procedure, taking $+$ for fermions and $-$ for bosons whenever $\pm$ is used. The first step of the thermofield approach is to purify the thermal state of the environment, $\hat{\rho}_{\beta}$. Renaming $\hat{a}_{\omega}\to\hat{a}_{1\omega}$, we introduce a new set of modes $\hat{a}^{\dag}_{2\omega},\hat{a}_{2\omega}$ that satisify the same (anti) commutation relations as $\hat{a}^{\dag}_{1\omega}\hat{a}_{1\omega}$ with hamiltonian 
$\hat{H}_{A} = \pm\int_{\omega_{1}}^{\omega_{2}}\omega\hat{a}^{\dag}_{2\omega}\hat{a}_{2\omega}$, but do not interact with the system or environment and thus will not affect the dynamics. In this enlarged space, we can define the thermofield double state $\ket{\Omega}$ such that the thermal expectation value of an operator $\hat{O}$ acting on the physical space is given by $\textrm{Tr}(\hat{\rho}_{\beta}\hat{O}) = \bra{\Omega}\hat{O}\ket{\Omega}$. We then introduce the following transformation for fermions
\begin{equation} \label{eq:B1}
\begin{pmatrix}
\hat{b}_{1\omega} \\
\hat{b}_{2\omega}
\end{pmatrix}
= 
\begin{pmatrix}
\textrm{cos}(\theta_{\omega}) & -\textrm{sin}(\theta_{\omega})  \\
\textrm{sin}(\theta_{\omega})  & \textrm{cos}(\theta_{\omega})  
\end{pmatrix}
\begin{pmatrix}
\hat{a}_{1\omega} \\
\hat{a}_{2\omega}
\end{pmatrix},
\end{equation}
and for bosons
\begin{equation}
\begin{pmatrix}
\hat{b}_{1\omega} \\
\hat{b}^{\dag}_{2\omega}
\end{pmatrix}
= 
\begin{pmatrix}
\textrm{cosh}(\theta_{\omega}) & -\textrm{sinh}(\theta_{\omega})  \\
-\textrm{sinh}(\theta_{\omega})  & \textrm{cosh}(\theta_{\omega})  
\end{pmatrix}
\begin{pmatrix}
\hat{a}_{1\omega} \\
\hat{a}^{\dag}_{2\omega}
\end{pmatrix},
\end{equation}
where $\theta_{\omega}$ is defined via $\textrm{sinh}(\theta_{\omega}) = \sqrt{ n_{\omega}}$ for bosons and $\textrm{sin}(\theta_{\omega}) = \sqrt{ n_{\omega}}$ for fermions where $n_{\omega} = 1/(e^{\beta(\omega-\mu)}\pm 1)$. The differences between the two transformations is to maintain the particle number conservation of the Hamiltonian for fermionic environments \cite{strachan2024extracting}. In this basis $\ket{\Omega}$ is the particle vacuum $\hat{b}_{1\omega}\ket{\Omega} = \hat{b}_{2\omega}\ket{\Omega}=0$ for bosons and a particle hole vacuum $\hat{b}_{1\omega}\ket{\Omega} = \hat{b}^{\dag}_{2\omega}\ket{\Omega}=0$ for fermions. The combined Hamiltonian is then given by 
\begin{align} \label{eq:B2}
\hat{H}_{A}+ \hat{H}_{E}+\hat{H}_{SE} &= \int_{\omega_{1}}^{\omega_{2}}d\omega\omega(\hat{b}^{\dag}_{1\omega}\hat{b}_{1\omega} \pm \hat{b}^{\dag}_{2\omega}\hat{b}_{2\omega}) \nonumber \\
&\qquad\qquad+\sum_{i}\int_{\omega_{1}}^{\omega_{2}}d\omega\sqrt{J_{i}(\omega)}(\hat{b}^{\dag}_{i\omega}\hat{Q}+\hat{Q}^{\dag}\hat{b}_{i\omega}),
\end{align}
where $J_{1}(\omega) = (1\mp n_{\omega})J(\omega)$ and $J_{2}(\omega) = n_{\omega}J(\omega)$. We discretize the environment using orthogonal polynomials to map the two star geometry environments to two one-dimensional tight-binding chains coupled to the system. To do this we introduce the following transformation
\begin{equation} \label{eq:B3}
\hat{b}_{i\omega} = \sum_{n=0}^{\infty}U_{in}(\omega)\hat{c}_{i,n},
\end{equation}
where $U_{in}(\omega) = \sqrt{J_{i}(\omega)}P_{in}(\omega)$ is defined via orthogonal polynomials $P_{in}(\omega)$, which obey the orthogonality relation
\begin{equation} \label{eq:B4}
\int_{\omega_{1}}^{\omega_{2}}d\omega J_{i}(\omega)P_{in}(\omega)P_{im}(\omega) = \delta_{nm}.
\end{equation}
Applying this transformation to the hamiltonian gives 
\begin{align} \label{eq:B5}
\hat{H}_{A}+ \hat{H}_{E}+\hat{H}_{SE} &= \sum_{i=1}^{2}\bigg[g_{i,0}(\hat{Q}^{\dag}\hat{c}_{i,0}+\textrm{h.c.})\nonumber \\
&\qquad\qquad+\sum_{n=0}^{\infty}\omega_{i,n}\hat{c}^{\dag}_{i,n}\hat{c}_{i,n} +(g_{i,n}\hat{c}^{\dag}_{i,n+1}+ \textrm{h.c.)}\bigg],
\end{align}
where $g_{i,0} = (\int_{\omega_{1}}^{\omega_{2}}d\omega J_{i}(\omega))^{\frac{1}{2}}$ and the coefficients $\omega_{i,n}$ and $g_{i,n\geq1}$ are defined through the recurrence relation 
\begin{equation} \label{eq:B6}
\pi_{i,n+1}(\omega) = (\omega-\omega_{i,n})\pi_{i,n}(\omega)-g_{i,n}\pi_{i,n-1}(\omega),
\end{equation}
where $\pi_{i,n}(\omega)$ are monic polynomials obtained by rescaling $P_{i,n}(\omega)$ such that the coefficient of the leading degree term is one. Overall, the star geometry Hamiltonian is mapped to two chains, each coupled to the system. Following recent works \cite{kohn2022quench,strachan2024extracting}, we use an interleaved ordering such that the environment particle hole pairs become nearest neighbours, decreasing entanglement growth. Orienting the system at the end of the MPS and interleaving the two chains as opposed to centering the system between two chains minimises the dimension of the interface the system has with the environment, which has also been shown to minimise entanglement generation~\cite{lacroix2025connectivitymattersimpactbath,kohn2022quench}.

\section{Parity Correction} \label{appendix: parity correction}
Here we show the correction needed for the evolution of odd parity fermionic operators, needed for calculating correlation functions like 
\begin{equation}\label{eq:C1}
G^{>}_{i}(t) = \textrm{Tr}(f_{i}(t)f_{i}^{\dag}\rho_{\infty}) = \textrm{Tr}(f_{i}(f_{i}^{\dag}\rho_{\infty})_{t}), 
\end{equation}
where $(...)_{t} = e^{-iHt}(...)e^{iHt}$. For the sake of clarity, for operators acting on $\mathcal{H}_{SE}$ we omit the ``\^{}" to distinguish them from operators just acting on the system or environment. The quantum regression theorem \cite{10.1093/acprof:oso/9780199213900.001.0001,gardiner2004quantum,carmichael2013statistical}, states that within the standard Markovian assumptions, the time evolution of operators of the form $\tilde{\rho}_{A}(t)$ is governed by the same Lindblad time evolution as $\hat{\rho}(t)$, provided the operator $A$ has the form $A=\hat{\mathbb{1}}_{B}\otimes\hat{X}_{S}$. In Breuer \cite{10.1093/acprof:oso/9780199213900.001.0001,BREUER200136}, they show that this also holds for the time convolutionless (TCL) equation but with respect to $\hat{\mathcal{L}}(t)$ and $\hat{\mathcal{I}}(t)$ rather than a fixed Lindblad operator. However, due to the fermionic anticommutation rules, $f^{\dag},f$ do not have this form. Despite this, as done in Ref.~\cite{PhysRevB.94.155142}, it is possible to rewrite Eq.~\eqref{eq:C1} in this way. We follow Ref.~\cite{PhysRevB.94.155142} and adopt the convention that product states in the combined Hilbert space  $\mathcal{H}_{SE}$ have the order $\ket{E}\otimes\ket{S}$, where $\ket{E}\in \mathcal{H}_E$ and $\ket{S}\in \mathcal{H}_S$. Defining $b_{i}$ and $f_{i}$ as environment and system operators respectively, we have $\{\hat{b}_{i},\hat{f}_{j}\} = 0$ which then gives
\begin{equation} \label{eq:C2}
f_{i} = (-1)^{\hat{N}_{E}}\otimes \hat{f}_{i},
\end{equation}
\begin{equation} \label{eq:C3}
b_{i} = \hat{b}_{i}\otimes \mathbb{1},
\end{equation}
where $N_{E}=\hat{N}_{E}\otimes\mathbb{1}$ is the number operator for the environment.
Now consider the greater Green's function, one of the two contributions to $G_{i}(t)$,
\begin{align} \label{eq:C4}
G^{>}_{i}(t) &= -i\textrm{Tr}\bigg(f_{i}(-1)^{N}\big((-1)^{N}f_{i}^{\dag}\rho\big)_t\bigg), \nonumber \\
&= -i\textrm{Tr}\bigg(F_{i}(-1)^{N_{S}}\big((-1)^{N_{S}}F_{i}^{\dag}\hat{\rho}\big)_t\bigg), \nonumber \\
&= -i\textrm{Tr}_{S}\bigg(\hat{f}_{i}(-1)^{\hat{N}_{S}}\textrm{Tr}_{E}\big((-1)^{N_{S}}F_{i}^{\dag}\hat{\rho}\big)_t\bigg),
\end{align}
where $N_S=\mathbb{1}_{B}\otimes\hat{N}_{S}$ is the number operator for the system and $[N,H] = 0$ has been used in the first line, and $F_{i} = (-1)^{N_{E}}f_{i} = \mathbb{1}_{B}\otimes \hat{f}_{i}$. This commutes with the environment operators $b_{\alpha}$, which means we can express $G^{>}_{i}(t)$ as 
\begin{equation} \label{eq:C5}
G^{>}_{i}(t) = \textrm{Tr}_{S}(\hat{f}_{i}\hat{\rho}_{f_{i}}(t)),
\end{equation}
where $\hat{\rho}_{f_{i}}(t) = (-1)^{\hat{N}_{S}}\textrm{Tr}_{E}\big((-1)^{N_{S}}F_{i}^{\dag}\rho\big)_t$, which satisfies the requirements for the time dependence inside the trace of $\hat{\rho}_{f_{i}}(t)$ to be given by the time convolutionless equation, as $(-1)^{N_{S}}F_{i}^{\dag} = \hat{\mathbb{1}}\otimes(-1)^{\hat{N}_{S}}\hat{f}_{i}^{\dag}$. Assuming we are on a timescale such that the system environment correlations have decayed away we can omit $\hat{\mathcal{I}}(t)$, which gives 
$\frac{d}{dt}\hat{\rho}_{f_{i}}(t)  = (-1)^{\hat{N}_{S}}\hat{\mathcal{L}}(t)\big((-1)^{N_{S}}F_{i}^{\dag}\hat{\rho}\big) = \underline{\hat{\mathcal{L}}}(t)\hat{\rho}_{f_{i}}(t)$, where $\underline{\hat{\mathcal{L}}}(t) = (-1)^{\hat{N}_{S}}\hat{\mathcal{L}}(t)(-1)^{N_{S}}$. This also holds for the lesser green's function $G^{<}_{i}(t) =\textrm{Tr}(f_{i}(\rho_{\infty}f_{i}^{\dag})_t)$, where $G_{i}(t) = -i(G^{>}_{i}(t)+G^{<}_{i}(t)).$

\section{Non-interacting Fermions} \label{appendix: non interacting fermions}

If $\hat{H}_{S}$ is quadratic in mode operators and particle number conserving, then $G^{<}_{i}(t) =\textrm{Tr}(\hat{f}^{\dag}_{i}\hat{f}_{i}(t)\hat{\rho}_{\infty})$ and $G^{>}_{i}(t) =\textrm{Tr}(\hat{f}_{i}(t)\hat{f}^{\dag}_{i}\hat{\rho}_{\infty})$ can be solved as single particle problems. Defining the system and environment mode operators together as $\{\hat{d}_{i}\}$, we have $\hat{H}=\sum_{ij}\boldsymbol{H}_{ij}\hat{d}^{\dag}_{i}\hat{d}_{j}$. Given $\hat{d}_{i}(t)=e^{i\hat{H}t}\hat{d}_{i}e^{-i\hat{H}t}$, we have 
\begin{align} \label{eq:D1}
\frac{\textrm{d}\hat{d}_{i}^{\dag}(t)}{\textrm{d}t} &= ie^{i\hat{H}t}[\hat{H},\hat{d}_{i}]e^{-i\hat{H}t}, \nonumber \\
&= -i\sum_{k}\boldsymbol{H}_{ik}\hat{d}_{k}(t),
\end{align} 
where we have used $[\hat{H},\hat{d}_{i}] = -\sum_{k}\boldsymbol{H}_{ik}\hat{d}_{k}$. Defining the vector of operators $\underline{d}(t) = (\hat{d}_{1}(t),...,\hat{d}_{N}(t))$ we have 
\begin{equation} \label{eq:D2}
\frac{\textrm{d}\underline{d}(t)}{\textrm{d}t} = -i\boldsymbol{H}\underline{d}(t),
\end{equation} 
which integrates to $\underline{d}(t) = e^{-i\boldsymbol{H}t}\underline{d}(0)$. Substituting this into the expressions for the greater and lesser greens functions gives 
\begin{equation} \label{eq:D3}
G^{<}_{i}(t) =\sum_{k}e^{-i\boldsymbol{H}_{ik}t}\textrm{Tr}(\hat{d}^{\dag}_{i}\hat{d}_{k}\hat{\rho}_{\infty}) = [\boldsymbol{U}(t)\boldsymbol{C}^{\rm SS}]_{ii},
\end{equation}
\begin{equation} \label{eq:D4}
G^{>}_{i}(t) =\sum_{k}e^{-i\boldsymbol{H}_{ik}t}\textrm{Tr}(\hat{d}_{k}\hat{d}^{\dag}_{i}\hat{\rho}_{\infty}) = [\boldsymbol{U}(t)(\mathbb{1}-\boldsymbol{C}^{\rm SS})]_{ii},
\end{equation}
where $\boldsymbol{C}^{\rm SS}_{ij} = \textrm{Tr}(\hat{d}_{j}^{\dag}\hat{d}_{i}\hat{\rho}_{\infty})$ is the equilbrium single particle correlation matrix.

\section{Nakajima-Zwanzig explanation} \label{appendix: NZ equation}
Here we provide an alternative discussion on the presence of a memory timescale $\tau_{\rm m}^{*}$ based on the Nakajima-Zwanzig equation. This can be expressed in two forms, given by the following 

\begin{align} \label{eq:E1}
\frac{\partial \hat{\rho}(t)}{\partial t}  &= i[\hat{\rho}(t),\hat{H}_{S}] + \int_{0}^{t}dt'\hat{\mathcal{K}}(t')[\hat{\rho}(t-t')], \\ 
&=i[\hat{\rho}(t),\hat{H}_{S}] + \int_{0}^{t-t_1}dt'\hat{\mathcal{K}}(t')[\hat{\rho}(t-t')] \nonumber \\ &\qquad\qquad+\textrm{Tr}_{B}(\hat{L}e^{(t-t_{1})\hat{Q}\hat{L}}\hat{Q}\hat{\rho}_{\textrm{tot}}(t_{1})). \label{eq:E2}
\end{align}
where $\hat{\mathcal{K}}(t')[\cdot] = \textrm{Tr}_{B}\big(\hat{L}e^{t'\hat{Q}\hat{L}}\hat{Q}\hat{L}\hat{P[\cdot]}\big),$  $\hat{P}[\cdot] = \textrm{Tr}_{B}[\cdot]\otimes\hat{\rho}_{B}$, $\hat{Q} = \mathbb{1}-\hat{P}$ and $\hat{L}[\cdot]=i[\cdot,\hat{H}]$. We now assume a unique steady state, $\lim_{t\to \infty}\hat{\rho}(t) = \hat{\rho}_{\infty}$, giving 
\begin{equation} \label{eq:E3}
0 = i[\hat{\rho}(t),\hat{H}_{S}] + \lim_{t\to\infty}\int_{0}^{t}dt'\hat{\mathcal{K}}(t')[\hat{\rho}(t-t')].
\end{equation}
\\
The RHS explicitly depends on $\hat{\rho}(0)$ if the memory kernel isn't bounded, so the steady state can only be unique if 
\begin{equation} \label{eq:E4}
||\int_{\tau^{*}_{m}}^{t}dt'\hat{\mathcal{K}}(t')[\hat{\rho}(t-t')]|| < \epsilon \:\:\:\forall \:\: t \geq \tau^{*}_{m},
\end{equation}
which then gives
\begin{equation} \label{eq:E5}
\frac{\partial \hat{\rho}(t)}{\partial t}  = i[\hat{\rho}(t),\hat{H}_{S}] + \int_{0}^{\tau^{*}_{m}}dt'\hat{\mathcal{K}}(t')[\hat{\rho}(t-t')] \:\:\:\forall \:\: t \geq \tau^{*}_{m}.
\end{equation}
Likewise, Eq.~\eqref{eq:E2} becomes
\begin{align} \label{eq:E6}
\frac{\partial \hat{\rho}(t)}{\partial t} &=i[\hat{\rho}(t),\hat{H}_{S}] + \int_{0}^{\tau^{*}_{m}}dt'\hat{\mathcal{K}}(t')[\hat{\rho}(t-t')] \nonumber \\ 
 &+\textrm{Tr}_{B}(\hat{L}e^{(t-t_{1})\hat{Q}\hat{L}}\hat{Q}\hat{\rho}_{\textrm{tot}}(t_{1})). \:\:\:\forall \:\: t-t_{1} \geq \tau^{*}_{m},
\end{align}
Since $t>0$, $t-t_{1} \geq \tau^{*}_{m}$ and thus $t >\tau^{*}_m$, meaning Eq.~\eqref{eq:E6} should be the same as Eq.~\eqref{eq:E5} which can only be true if
\begin{equation} \label{eq:E7}
||\textrm{Tr}_{B}(\hat{L}e^{(t-t_{1})\hat{Q}\hat{L}}\hat{Q}\hat{\rho}_{\textrm{tot}}(t_{1}))|| < \epsilon \:\:\:\forall \:\: t-t_{1} \geq \tau^{*}_{m}.
\end{equation}
Physically, this means the effect of system-environment correlations at time $t$ is also lost in the time $\tau^{*}_{m}$. 

\section{Thermalisation for the SIAM} \label{appendix:SIAM thermalisation}

\begin{figure}[h!] % t or b to control placement
  \centering
    \includegraphics[width=0.8\linewidth]{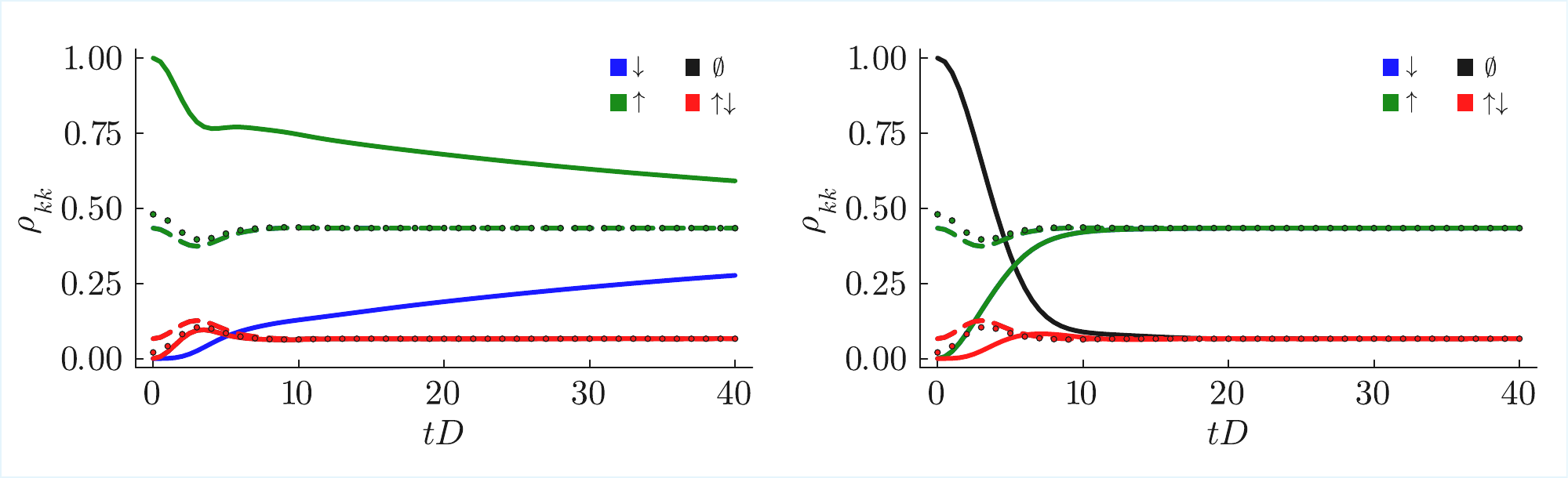}
    \caption{Thermalisation of the diagonal elements $\hat{\rho}_{kk}$, $k=\{\emptyset,\uparrow,\downarrow,\uparrow\downarrow\}$ for various initial states for the $\beta=20/D$ SIAM case. The dotted lines represent $\hat{\rho}(0)=\hat{\rho}_{f}$, the dashed lines represent $\hat{\rho}(0) = \hat{\rho}_{\infty}$ and the solid lines represent (a) $\hat{\rho}(0) =  \ket{\uparrow}\bra{\uparrow}$ and (b) $\hat{\rho}(0) =  \ket{\emptyset}\bra{\emptyset}$. (a)Thermalisation for the initial states $\hat{\rho}_{f}$, $\hat{\rho}_{\infty}$ and the spin up state $\ket{\uparrow}\bra{\uparrow}$.(b) Thermalisation  for the initial states $\hat{\rho}_{f}$, $\hat{\rho}_{\infty}$ and the empty state $\ket{\emptyset}\bra{\emptyset}$. Note that due to symmetry, $\hat{\rho}_{\uparrow\uparrow} = \hat{\rho}_{\downarrow\downarrow}$ if they are the same initially, and the same is true for $\hat{\rho}_{\uparrow\downarrow,\uparrow\downarrow}$ and $\hat{\rho}_{\emptyset,\emptyset}$. Parameters: $\Gamma=\pi D/20,\;U=5\Gamma,\;\mu=0$, $N_{B}=80$, maximum bond dimension $\chi_{\textrm{max}}=150$ (same as ~\cite{PhysRevB.104.014303}).}
  \label{fig:SIAM thermalisation}
\end{figure}

Here we discuss the thermalisation of the SIAM, specifically the $\beta= 20/D$ case but the $\beta=2.5/D$ is similar. Fig.~\ref{fig:SIAM thermalisation}(a) shows the thermalisation in terms of the diagonal elements of the state for the three initial states $\hat{\rho}_{f}$, $\hat{\rho}_{\infty}$ and the spin up state $\ket{\uparrow}\bra{\uparrow}$. It's clear that there is little difference in the thermalisation timescales of $\hat{\rho}_{f}$ and $\hat{\rho}_{\infty}$, meaning no significant extreme NMQMpE exists for this case. However, when comparing to the thermalisation of the spin up state, the difference is clear, with the spin up state thermalising significantly more slowly. This demonstates how our method can provide significant thermalisation even in the absence of an extreme NMQMpE~\cite{PhysRevLett.134.220403}, as $\hat{\rho}_{f}$ by construction provides the fastest path to thermalisation when starting from a product state. If there is no extreme NMQMpE, then that simply means $\hat{\rho}_{\infty}$ thermalises just as quickly, it does not mean that this pathway is not fast, as can be seen by comparing to the slow thermalisation of $\hat{\rho}(0)=\ket{\uparrow}\bra{\uparrow}$. In the case of the SIAM, there is a simple canonical choice for the initial state that does thermalise on a similar scale to $\hat{\rho}_{f}$ and $\hat{\rho}_{\infty}$, namely $\hat{\rho}(0)=\ket{\emptyset}\bra{\emptyset}$~\footnote{By symmetry, $\hat{\rho}(0)=\ket{\uparrow\downarrow}\bra{\uparrow\downarrow}$ gives rise to the same dynamics.}. This is shown in Fig.~\ref{fig:SIAM thermalisation}(b) where all three states relax on roughly the same timescale. However, in less simple models it may be less clear if there exists a simple initial state that thermalises as fast as $\hat{\rho}_{f}$, but through our method we always use the initial state that thermalises most quickly, without any additional analysis needed.

\bibliography{References.bib}

\end{document}